%% file: main.tex
\documentclass[twocolumn,lengthcheck,superscriptaddress]{revtex4-1}

\usepackage[english]{babel}
\usepackage[T1]{fontenc}
\usepackage{amsmath}
\usepackage{amsfonts}
\usepackage{hyperref}
\usepackage{braket}
\usepackage{amssymb}
\usepackage{graphicx}
\usepackage{hyperref}
\usepackage{color, caption}
\usepackage{comment}
\usepackage{tikz}
\usepackage{tikzit}
\input{Tensors.tikzstyles}
\usetikzlibrary{calc}

\usetikzlibrary{positioning,shapes.geometric,decorations.markings,arrows,knots,hobby} 
\usepackage[many]{tcolorbox}
\usepackage{pgfplots}
\usepgfplotslibrary{patchplots}
\usepgfplotslibrary{fillbetween}
\usepackage{dsfont}

\usepackage{caption}
\usepackage{subcaption}

\pgfplotsset{compat=1.18}

\begin{document}

\title{Linear-nonlinear 
duality for circuit design on quantum computing platforms
}

\author{William E. Salazar}
\affiliation{Centre for Bioinformatics and Photonics---CIBioFi, Edificio E20 No.~1069, Universidad del Valle, 760032 Cali, Colombia}
\affiliation{Departamento de Física, Universidad del Valle, 760032 Cali, Colombia}

\author{Omar Calderón-Losada}
\email{Electronic Address: omar.calderon@correounivalle.edu.co}
\affiliation{Centre for Bioinformatics and Photonics---CIBioFi, Edificio E20 No.~1069, Universidad del Valle, 760032 Cali, Colombia}
\affiliation{Departamento de Física, Universidad del Valle, 760032 Cali, Colombia}

\author{John H. Reina}
\email{Electronic Address: john.reina@correounivalle.edu.co}
\affiliation{Centre for Bioinformatics and Photonics---CIBioFi, Edificio E20 No.~1069, Universidad del Valle, 760032 Cali, Colombia}
\affiliation{Departamento de Física, Universidad del Valle, 760032 Cali, Colombia}

\begin{abstract}

Beam splitters (BSs) and optical parametric amplifiers (OPAs) can be described using Lie groups $SU(2)$ and $SU(1,1)$. Here, we show that the dynamical trajectories of these devices are connected via a Wick rotation on their respective group manifolds. This yields an exact amplitude-level duality between BSs of transmittance $\eta$ and OPAs of gain $g=1/\eta$. This geometric correspondence admits a compact tensor-network formulation, which we use to construct a circuit-model protocol that reproduces PDC transition amplitudes. This construction naturally leads  to finite-dimensional, truncated PDC unitaries that exactly reproduce the first $q$ amplitudes of an ideal parametric amplifier. Our results demonstrate that key amplitude-level features of nonlinear optical processes can be simulated using only native single-qubit unitaries and measurement-based primitives on existing digital quantum hardware. This  extends PDC-inspired entanglement-generation mechanisms beyond photonic architectures.
\end{abstract}

\maketitle
\section{Introduction}

Quantum technologies, and in particular quantum computing, have experienced a remarkable surge in recent years. Their potential to transform computing, communication, and sensing has generated significant attention~\cite{Preskill/Nisq/2018,Biamonte/Q_machine_learnign/2017,pffaf/Quantum_teleportation/2018,Omar/QSensing23,Giovanneti/Quantum_measurents/2004,JH/Q_Engines23}. Among the various quantum-technological platforms currently under active development, photonic architectures stand out due to their high fidelities, scalability, and low error rates~\cite{Jacques/universal_optics/2019,Flamini/photonic_review/2019,Zhong/computational_advantage_photons/2019,PhotonQP22}. These characteristics have positioned photonics at the forefront of several demonstrations of quantum advantage.

Within any optical-based quantum platform, two devices are indispensable: beam splitters and optical parametric amplifiers (OPAs). Beam splitters enable the creation of coherent superpositions, while OPAs generate entangled photon pairs via squeezing~\cite{Hong/original_HOM/1987,Mandel/PDC_experiments/1995,Bouchard/HOM/2021,Omar/Entanglement_experiments22,Milburn/Quantum_optics}. Modern quantum technologies, especially quantum computing, have demonstrated how both types of resources --linear and nonlinear-- can be exploited in a unified manner to achieve quantum advantage~\cite{Preskill/quantum_advantage/2012,Harrow/Computational_supremacy/2017,Eisert/quantum_random_sampling_advantage/2023,Zhong/computational_advantage_photons/2019,Neil/Super_conducting_supremacy/2018,Wang/Boson_samping_advantage/2023,JH/Qadvantage17}. Despite this conceptual parity, their physical realizations differ substantially: a lossless beam splitter is a passive, number-conserving device typically implemented using simple crystals, whereas an OPA relies on a nonlinear crystal that does not conserve photon number. This fundamental distinction motivates a deeper examination of how similar functionality might --or might not-- translate across different quantum platforms.

In non-photonic quantum architectures~\cite{Georgescu/ion_trap_rev/2020,Chatterjee/super_conductin_practice/2021,JH/QDs03,Henriet/neutral_atom_QC/2020,Auger/rydberg_blueprint/2017,JH/spins04,Gonzalez/silicon_based_review/2021}, the underlying physical mechanisms differ drastically from those in optics. Nevertheless, the role of a 50:50 lossless beam splitter can be faithfully mapped to that of a Hadamard gate. More generally, in the KLM scheme for linear optical quantum computing~\cite{KML/Quantum_computing/2001,Milburn/Review/2007}, $Y$-rotation unitaries correspond one-to-one with beam splitters of arbitrary transmittance. Given this tight correspondence, it is natural to ask whether a similar identification exists for the nonlinear counterpart: is there a device or operation in non-photonic platforms that could effectively mimic a parametric amplifier and its associated family of unitaries?

Identifying an analog of parametric amplifiers in non-photonic (digital) platforms presents significant challenges. Although Trotterization offers a universal approach for simulating dynamics~\cite{Lloyd/universal_simulators/1996,Childs/Trotter_error/2021}, directly encoding the generator of the parametric down-conversion (PDC) process into a qubit-based system is problematic because the effective generator involves non-number-conserving operators. Naively embedding such a generator would lead to non-unitary dynamics, making a direct implementation infeasible. Moreover, a Trotter-type decomposition of the corresponding exponential becomes highly inefficient in the high-gain regime~\cite{Su/Product_formulas/2019,Childs/Trotter_error/2021}.

In addition to Trotter-type methods, more advanced Hamiltonian-simulation techniques such as qubitization and quantum signal processing offer asymptotically optimal scaling for a broad class of dynamics~\cite{Low/Chuang/Qubitization/2019,Low/QSP/2017}. However, their applicability relies on the ability to construct efficient  block-encodings of the target generator. For the PDC process, whose effective 
generator is non-number-conserving and whose strength grows rapidly in the high-gain regime, such block-encodings are not straightforward to obtain. 
Consequently, even optimal Hamiltonian-simulation frameworks do not provide a direct or efficient route to implementing PDC-like transformations within qubit-based platforms.

Motivated by these difficulties, we turn to more fundamental considerations regarding how photonic and non-photonic platforms employ their respective resources. In photonic systems, OPAs enable the well-known PDC nonlinear process, yet direct analogues of such nonlinear interactions are far less common in other platforms. Rather than attempting to replicate the nonlinear interaction itself, here we adopt a different perspective: we propose an alternative route that bypasses the need for directly mapping nonlinear processes to PDC. This perspective allows us to extend the conceptual notion of parametric amplification to quantum platforms that do not rely on photonic interactions.

We develop a physically motivated argument relating the counting statistics of an OPA (a nonlinear device) to those of a lossless beam splitter (a linear device). Given the standard role of beam splitters as rotation gates in various quantum computing schemes~\cite{KML/Quantum_computing/2001,Milburn/Review/2007}, our goal is to determine whether an operationally meaningful analogue of a parametric amplifier can be defined and implemented within integrated qubit systems. In other words, we ask whether the essential statistical features of PDC can be abstracted and exported to non-photonic computational settings.

The structure of this work is as follows. In Sec.~\ref{sec: Mathematical_argument}, we establish a connection--grounded in their common Lie-group structure--between the unitaries of a two-mode, lossless beam splitter with transmittance $\eta$ and a two-mode, lossless parametric amplifier with gain $g$. This connection relies on a physically intuitive argument that exploits the $SU(2)$ and $SU(1,1)$ group structures associated with the BS and PDC, respectively. In Sec.~\ref{sec: Diagramatic_intuition}, we introduce a circuit-model protocol that simulates the action of a PDC in terms of beam-splitter operations at the level of matrix elements. The protocol encodes bosonic occupation numbers and uses a suitable teleportation mechanism to swap the photon number entering one of the modes. Building on this protocol, in Sec.~\ref{sec: pdc_concept}, we define  the notion of a parametric amplifier (up to $q$)-gate: a unitary capable of simulating the generation of entangled photon pairs--binary encoded in qubits--up to order $q$. This framework extends PDC-based entanglement generation to non-photonic quantum platforms. Finally, Sec.~\ref{sec:conclusions_outlook} summarizes our main results and discusses future research directions.

\section{Beam splitters as Parametric-Down converters in Euclidean time}
\label{sec: Mathematical_argument}

\begin{figure}
   \centering
\includegraphics[width=8cm]{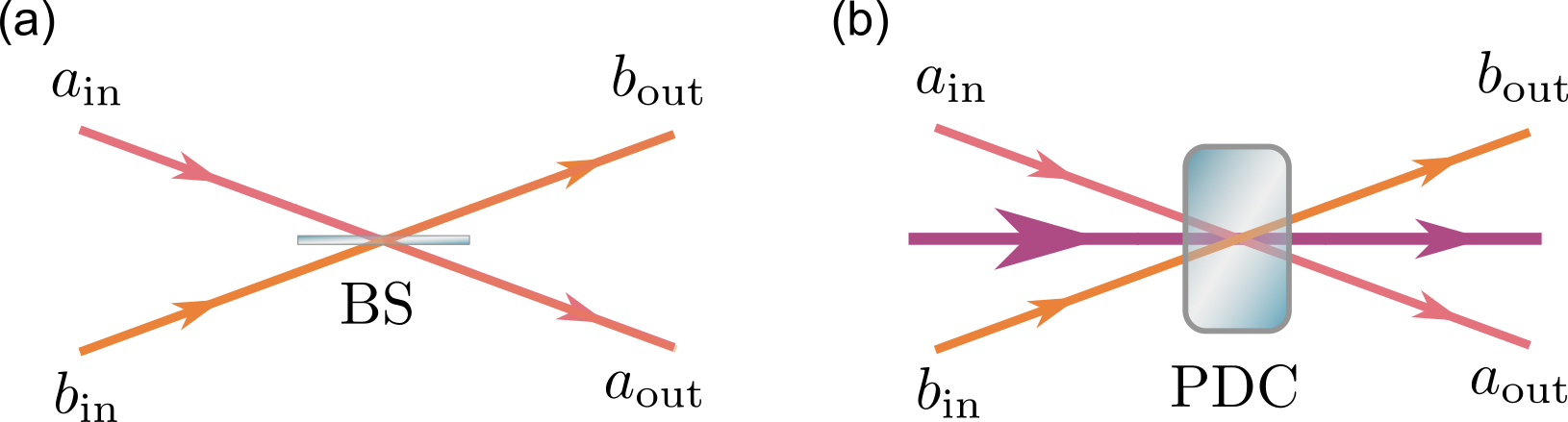}
\caption{Schematics of a (a) two-mode beam splitter (BS) and an (b) optical parametric amplifier (PDC). The purple beam indicates the pump beam undergoing down-conversion. The bosonic operators of the input light beams are rotated by the action of each optical device as dictated by the unitaries in the equations~\eqref{eqn:BS_rotation} and~\eqref{eqn:PDC_rotation}, respectively. For a BS of transmittance $\eta$, the input modes are rotated by a $SU(2)$ unitary, while for a PDC of gain $g$, the input modes are rotated by a $SU(1,1)$ unitary that does not preserve the total number of photons entering the optical device.}
\label{fig:BS_PDC_diagram}
\end{figure}

We begin by establishing some basic results of the representation theory of the Lie groups $SU(2)$ and $SU(1,1)$. In particular, we focus on their interpretation in the context of optical devices.
To set the grounds, let $a$ and $b$ denote the annihilation operators of the bosonic modes (light beams) entering a two-mode optical device (see Fig.~\ref{fig:BS_PDC_diagram}). At the operator level, the action of the particular optical device is to rotate the two modes, and for a BS of transmission $\eta$,
\begin{align}
\label{eqn:BS_rotation}
 \begin{pmatrix} a_{out} \\ b_{out} \end{pmatrix} \;
 \to \begin{pmatrix} \cos(\theta) & \sin(\theta) \\ -\sin(\theta) & \cos(\theta) \end{pmatrix} 
 \begin{pmatrix} a \\ b \end{pmatrix},
\end{align}
with $\cos^{2}\theta =\eta$. On the other hand, for a PA of gain $g$ one yields,
\begin{align}
\label{eqn:PDC_rotation}
 \begin{pmatrix} a_{out} \\ b_{out} \end{pmatrix} \;
 \to \begin{pmatrix} \cosh(\phi) & \sinh(\phi) \\ \sinh(\phi) & \cosh(\phi) \end{pmatrix} 
 \begin{pmatrix} a \\ b \end{pmatrix}
\end{align}
with $\cosh^{2}(\phi) =g$.

The two-mode bosonic algebra can be used to construct realizations of the $\mathfrak{su}(2)$ and $\mathfrak{su}(1,1)$ algebras via the Schwinger mapping~\cite{Ruan/Two_boson_Higgs_Algebra/2003,sakurai/Modern/2003} (see Appendix A). In terms of $\{J_{x},J_{y},J_{z}\}$ and $\{K_{x},K_{y},K_{z}\}$, the generators for $\mathfrak{su}(2)$ and $\mathfrak{su}(1,1)$ respectively, the unitaries $U_{\text{BS}}^{\eta}$ describing a BS with transmittance $\eta$, and $U_{\text{PDC}}^{g}$ describing a parametric amplifier of gain $g$, are respectively given by the following one-parameter family of curves in
the $SU(2)$, and $SU(1,1)$ group manifolds,
\begin{equation}
 \label{eq:BS_PDC_unitaries}
 U_{\text{BS}}^{\eta} = e^{2i \theta J_{y}}\, , \quad U_{\text{PDC}}^{g} = e^{2i \phi K_{y}}.
\end{equation}

\begin{figure}
    \centering
\includegraphics[width=8.3cm]{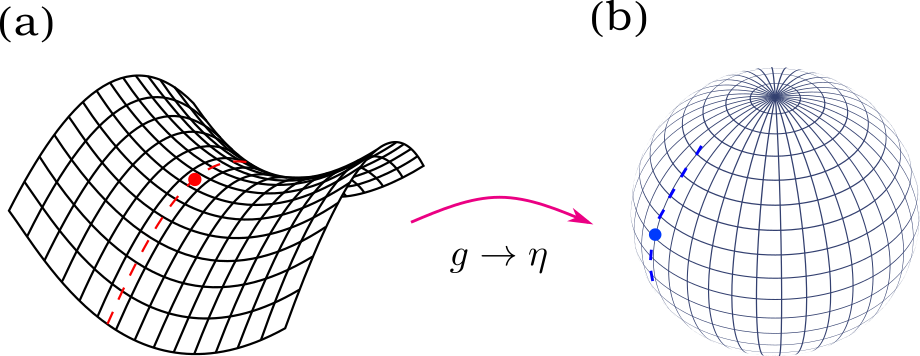}
\caption{Beamsplitter as a Euclidean parametric amplifier. A PDC of gain $g$ is represented by a point on a one-parameter curve on the group manifold $SU(1,1)$. Locally, $SU(1,1) \cong SO(1,2)$, and the parametric amplifier acts transitively on the two-dimensional hyperbolic space (as represented in panel (a)). After the Wick rotation $g\to\eta$, the one-parameter curve in $SO(1,2)$ transforms into a curve in $SO(3)$ (locally, $SU(2)$), which acts transitively on the two-dimensional sphere (panel (b)). This identifies each beam splitter as a Euclidean parametric amplifier with ``inverse'' temperature $\eta = g^{-1}$.} 
    \label{fig: geometric_picture}
\end{figure}

The group-theoretic interpretation of optical devices has been used in quantum optics, especially in the description of $SU(1,1)$ vs. $SU(2)$ interferometers~\cite{Klauder/su2_interferometry/1986, Jing/2011, Chekhova:16, Ou/Li/2020}. On the algebraic level, $\mathfrak{su}(2)$ and $\mathfrak{su}(1,1)$ have the same complexification in $\mathfrak{sl}(2,C)$, which means that their representation theory is identical. However, as groups, $SU(1,1)$ is non-compact while $SU(2)$ is compact. This difference between the dynamical groups of the two optical devices can be understood in terms of their respective group manifolds, since there is a local isomorphism between $SU(1,1)$ and the group of Lorentz transformations in 2+1 dimensions $SO(1,2)$. This means that the unitary of a parametric amplifier of gain $g$ can be locally identified as a
one-parameter curve in the $SO(1,2)$ manifold {(see Fig.~\ref{fig: geometric_picture}(a))}. After an imaginary time rotation, $SO(1,2)$ transforms into $SO(3)$ {(see Fig.~\ref{fig: geometric_picture}(b))}, which is the isometry group of 3-dimensional Euclidean space. At the local level, $SO(3)$ is again isomorphic to $SU(2)$, which is the dynamical group of beam splitters. Within this interpretation, both $U_{\text{BS}}^\eta$ and $U_{\text{PDC}}^g$ are dual to each other via a Wick rotation. The Euclidean time rotation leads to the exact relation
\begin{equation}
\label{eqn:duality_matrix_elements}
 \braket{l, s\left|U_{\text{PDC}}^{g}\right|n, m}=\frac{1}{\sqrt{g}}\braket{l, m|U_{\text{BS}}^{1/g}|n, s},
\end{equation}
between the matrix elements of both unitaries (see Appendix~\ref{appendix:Lie-Group-structure}).
We note that while the previous connection between the BS and PDC transition amplitudes was presented in~\cite{Cerf/Two_boson_interference/2019}, we now propose that this duality arises naturally from the geometry of both $SU(1,1)$ and $SU(2)$ Lie groups.
 
Returning to the concept of duality, and since the transmittance of a lossless beam splitter ranges within $0 \leq \eta = g^{-1} \leq 1$, the relationship between matrix elements in Eq.~\eqref{eqn:duality_matrix_elements} implies that the duality only relates the action of beam splitters of arbitrary transmittance with parametric amplifiers in the high gain regime, $g \geq 1$.
\section{Diagrammatic interpretation and composite encoding}
\label{sec: Diagramatic_intuition}

\begin{figure}
	\centering
\includegraphics[width=8.5cm]{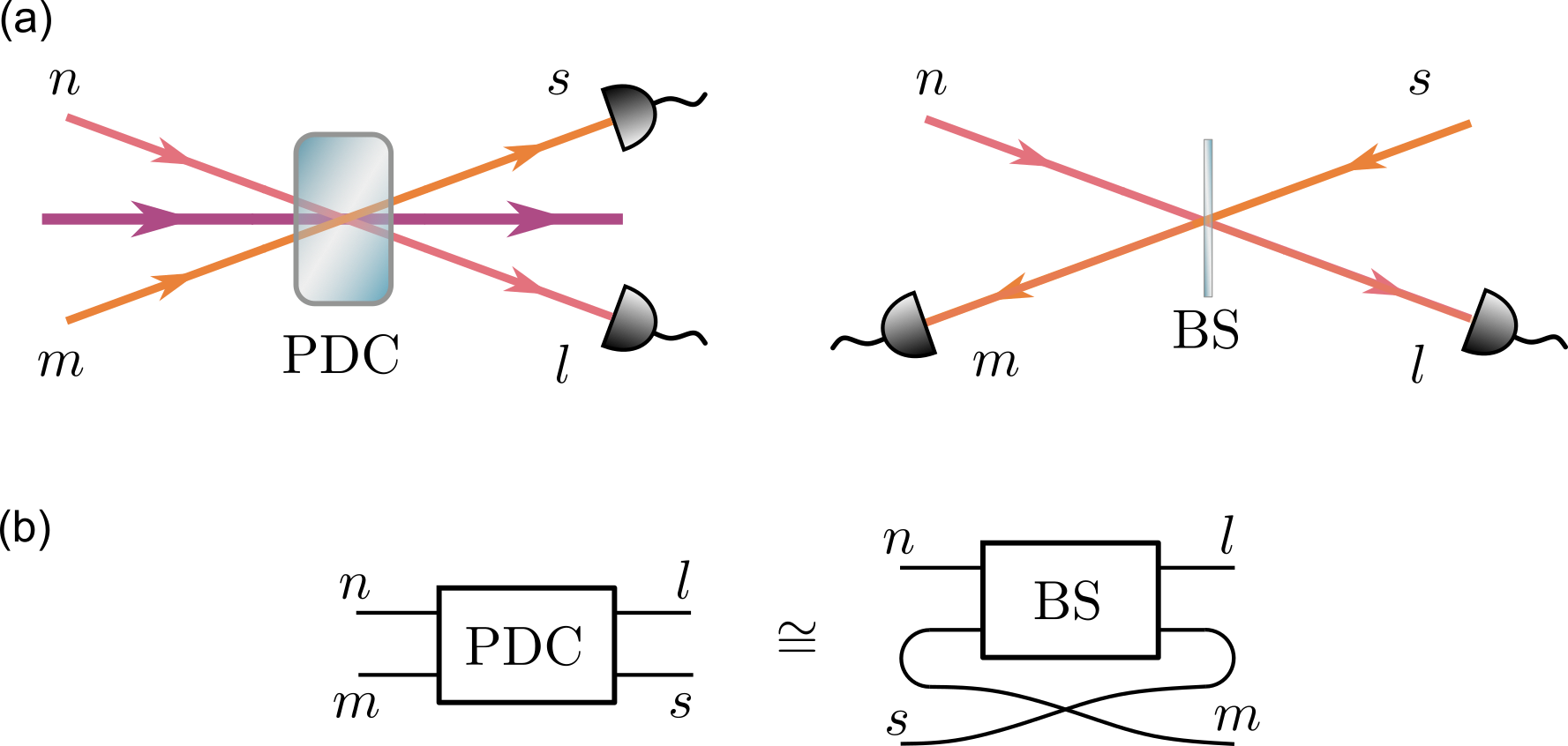}
	\caption{BS--PDC duality: (a) The probability amplitude of detecting $s$ and $l$ photons at the output of a parametric amplifier with gain $g$, given input photons $n$ and $m$, is dual to the probability amplitude of detecting $m,l$ photons, given $n,s$ entering a lossless beam splitter with transmittance $1/g$. (b) At the level of matrix elements, this equivalence requires swapping the photon numbers entering and leaving the lower mode. The equality in panel (b) must be interpreted modulo the global factor $1/\sqrt{g}$.}
\label{fig:pre-tensor_network}
\end{figure}

The relationship between the BS and PDC one-parameter families of curves shown in the previous section (arising from a Wick rotation of the corresponding Lie-algebra generators) suggests the possibility of reproducing the action of one optical device in terms of the other, at least at the level of matrix elements. In particular, Eq.~\eqref{eqn:duality_matrix_elements} shows that, after exchanging the number of photons injected in one of the modes, a beam splitter becomes a parametric amplifier with reciprocal transmittance. This observation is illustrated in Fig.~\ref{fig:pre-tensor_network}(a). 

In a tensor-network language, this duality can be represented diagrammatically, as shown in Fig.~\ref{fig:pre-tensor_network}(b). The PDC box (left) represents a two-mode parametric amplifier of gain $g$, while the BS box (right) denotes a lossless beam splitter with transmittance $1/g$. Each wire corresponds to an optical mode, and the integer labels indicate the number of photons either injected ($n,m$ and $n,s$ on the left wires) or measured ($l,s$ and $l,m$ on the right wires). As indicated, the equivalence in Fig.~\ref{fig:pre-tensor_network}(b) must be interpreted modulo the global factor $1/\sqrt{g}$ appearing in Eq.~\eqref{eqn:duality_matrix_elements}.

The concrete realization of the duality in Eq.~\eqref{eqn:duality_matrix_elements} requires exchanging the photon-number label of the lower mode. Physically, such a swap operation on occupation numbers violates causality: one would need to condition on measurement outcomes before the input state is prepared. This is, of course, impossible at the physical level. 

However, this apparent obstruction disappears once the occupation numbers are encoded into quantum registers. Specifically, we introduce a binary encoding $E_{B}: \mathbb{N}\to\{0,1\}^{Q}$, where $Q$ denotes the required number of qubits. After encoding, the problematic SWAP in Fig.~\ref{fig:pre-tensor_network}(b) can be replaced by a teleportation-like operation between the encoded qubits and a set of ancilla qubits in the diagrammatic (categorical) sense~\cite{Coecke/ZX/2023}. This effectively reroutes the photon-number labels without violating causality. The complete circuit structure implementing this idea is shown in Fig.~\ref{fig:Circuit_scheme}.
\begin{figure}
\ctikzfig{Circuit_scheme_v2}
    \caption{Circuit-model scheme for the composite encoding in Fig.~\ref{fig:full_encoding}. 
    Different input states contribute different numbers of qubits to the binary encoding. 
    The color code blue/red indicates the qubits associated with the photon numbers entering the second/first mode.
    To ensure that the composite encoding extends to a unitary map on all computational states, we introduce ancilla qubits that remain unaffected by the optical transformation.
    To exchange the number of occupants in the lower mode, two additional quantum registers of $\lceil\log m\rceil$ qubits each are required. 
    These registers are prepared in $\lceil\log m\rceil$ EPR pairs, represented by cups at the start of the circuit, and subsequently measured in the Bell basis, represented by caps, following the standard categorical teleportation structure~\cite{Biamonte/tensor_networks/2019}.}
\label{fig:Circuit_scheme}
\end{figure}

The occupation-number representation serves merely as shorthand for vectors in the symmetric sector of the multiparticle Hilbert space. A SWAP operation cannot act directly on these occupation-number states. Instead, we encode each classical integer into a binary qubit string using the canonical encoding
\begin{equation}
\ket{n}\otimes\ket{m} \mapsto
\ket{x_{N-1}x_{N-2}\ldots x_0} \otimes
\ket{y_{M-1}y_{M-2}\ldots y_0},
\end{equation}
where $N=\lceil\log_2(n+1)\rceil$ and $M=\lceil\log_2(m+1)\rceil$. 
The total number of qubits required for binary encoding grows as 
$\mathcal{O}(\log_{2}(nm))$. 
For states containing only a single boson per mode, this encoding is optimal, as the number of qubits equals the number of modes.

The action of an optical device cannot be applied directly to the binary-encoded states; it acts naturally only on physically symmetrized states. For example, consider the occupation-number state $\ket{1,1}$ and denote by $\ket{a}$ and $\ket{b}$ the single-particle states entering the first and second modes, respectively. 
For a BS, the action on each particle is equivalent to a rotation $R_y(\theta)$:
\begin{equation}\label{eqn: BS_ry_rotation}
U^{\eta}_{BS}\ket{1,1} = R_{y}(\theta)\otimes R_{y}(\theta)\,\ket{\psi^{+}},
\end{equation}
with $\ket{\psi^{+}}=(\ket{ab}+\ket{ba})/{\sqrt{2}}$. 
We denote the map taking arbitrary tensor-product states into symmetrized states by $E_{P}$, the \emph{physical encoding}. The binary-encoded states must be mapped through the composite encoding 
$E_{P}\circ E_{B}^{-1}$
prior to entering the optical device (see Fig.~\ref{fig:full_encoding}). 

The distinction between binary and physical encoding becomes essential in the second part of the protocol: occupation numbers are first encoded to perform the swap operation at the computational level, and then symmetrized to apply the beam-splitter transformation. Although we employ the canonical binary encoding, it may be replaced by alternative encodings optimized for resource efficiency~\cite{Moffat/Binary_Encoding/2008}.
\begin{figure}
\ctikzfig{Coding}
\caption{Composite encoding of occupation-number states into physical qubit states. Binary-encoded states are transformed into symmetric states by the composite map $E_{P}\circ E_{B}^{-1}: \mathcal{H}^{\mathcal{O}(\log_{2}(nm))}\to \mathcal{S}(\mathcal{H}^{\otimes (n+m)})$.}
\label{fig:full_encoding}
\end{figure}

\section{The q-PDC concept}\label{sec: pdc_concept}

There is a fundamental constraint for simulating a parametric amplifier within a circuit model: its non–particle-conserving (active) character. For an arbitrary initial state with $n$ and $m$ photons entering the first and second modes, respectively, the action of a lossless parametric amplifier produces a superposition of states with a fixed photon-number imbalance, i.e. $n-m$ is conserved. This implies that one cannot directly construct a finite-dimensional unitary $U_{\text{PDC}}$ that reproduces the exact matrix elements of a PDC for arbitrary two-mode photonic input states. The reason is simple: since $U_{\text{PDC}}$ does not preserve the total number of photons, any faithful representation must act on an infinite-dimensional Hilbert space.

\begin{figure}
    \centering
\includegraphics[width=8cm]{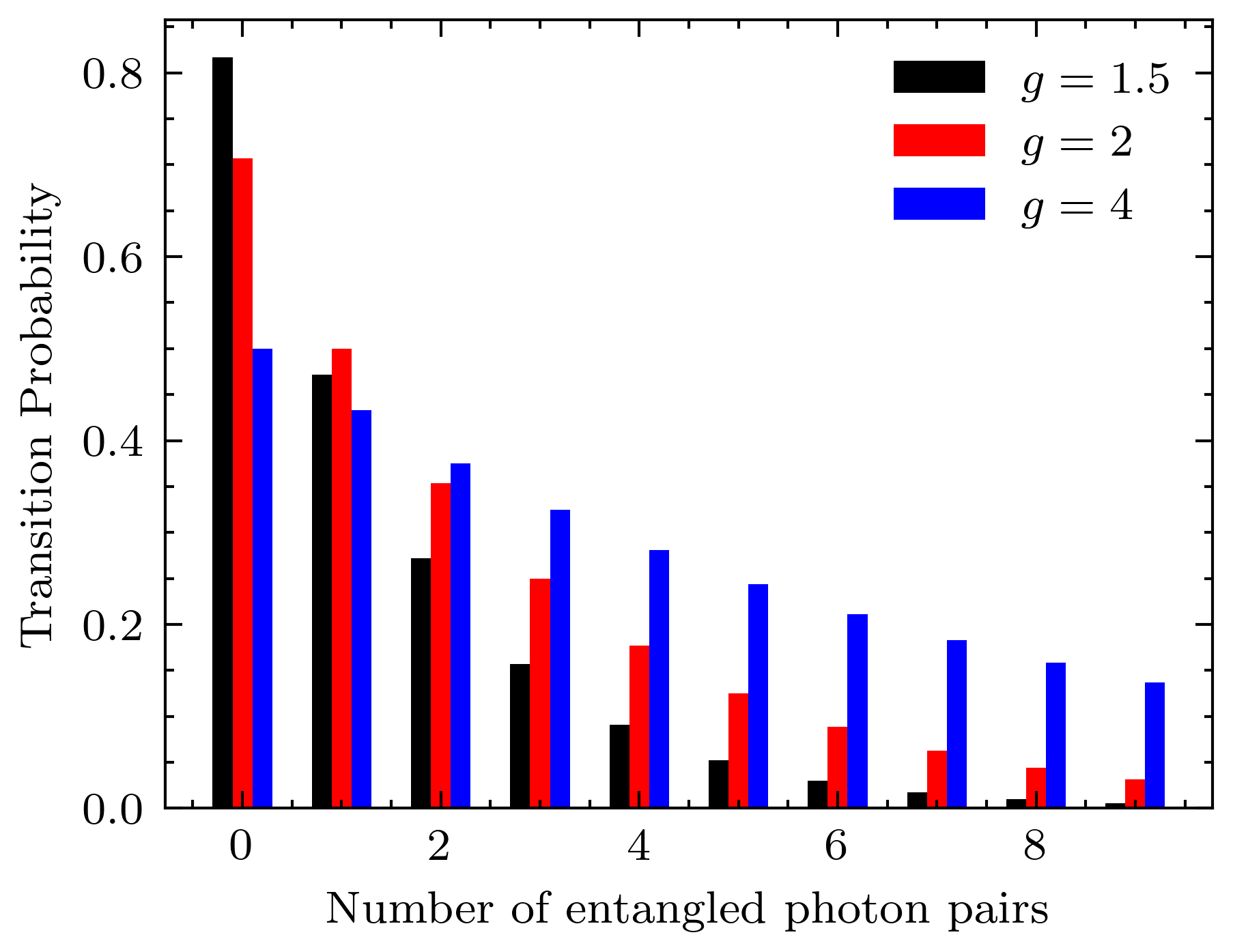}
\caption{Probability of pair creation of multi-photon entangled states for a parametric down-conversion device in the high-gain $g>1$ regime. For any fixed gain, the probability of multi-photon creation from the vacuum decreases with the number of entangled photons. As the parametric gain increases, the probability distribution becomes broader.}
\label{fig:pair_photonic_decay}
\end{figure}

From a physical viewpoint, however, one is often not interested in all possible matrix elements, but rather in the pair production of entangled photons from the vacuum, namely
\begin{equation}
U_{\text{PDC}}^{g}\ket{0,0}=\sum_{l \in \mathbb{N}}c_{l}(g)\ket{l,l},
\end{equation}
where the transition amplitudes are given by $c_{l}(g)=\tanh{\left(\cosh^{-1}{\sqrt{g}}\right)}^{l}/{\sqrt{g}}$, and their profile is shown in Fig.~\ref{fig:pair_photonic_decay}. For fixed gain $g$, the probability of creating multi-photon entangled pairs decreases rapidly with the number of photons. Motivated by this decay, we introduce the \emph{parametric amplifier up to order $q$} gate, $U_{\text{PDC},q}^{g}$, as the unitary that reproduces the vacuum transition amplitudes of a lossless parametric amplifier of gain $g$ up to $q$ entangled photon pairs:
\begin{equation}
\label{eq:pdc_up_to_q}
U_{\text{PDC},q}^{g}\ket{0,0} = \sum_{l=0}^{q}\tilde{c}_{l}(g)\ket{l,l},
\end{equation}
with $|\tilde{c}_{l}(g)|^{2}=|c_{l}(g)|^{2}$. By construction, this defines a hierarchy of $q$-parametric amplifiers,
\begin{equation}
U_{\text{PDC},1}^{g} \subseteq U_{\text{PDC},2}^{g} \subseteq U_{\text{PDC},3}^{g} \subseteq \dots \subseteq U_{\text{PDC},q}^{g},
\end{equation}
where the inclusion symbol is meant in the sense that $U_{\text{PDC},q}^{g}$ extends the action of $U_{\text{PDC},q-1}^{g}$ onto a larger subspace of the Fock space. In the formal limit $q\to \infty$, one recovers the full infinite-dimensional unitary representation $U^{g}_{\text{PDC}}$.

This definition can be generalized to arbitrary initial states with non-vanishing imbalance. For an input state $\ket{n,m}$ with fixed $n-m$, we define
\begin{equation}
 U_{\text{PDC},q}^{g}\ket{n,m} = \sum_{l=0}^{q}\tilde{c}_{l}(g)\ket{n-m + l,l},
\end{equation}
where $|\tilde{c}_{l}(g)|^{2}=|\braket{n-m+l,l|U_{\text{PDC}}^{g}|n,m}|^{2}$. In this way, $U_{\text{PDC},q}^{g}$ reproduces the exact PDC transition probabilities within the truncated subspace defined by a fixed imbalance and a maximum pair number $q$.

\begin{figure} 
    \centering
\includegraphics[width=8.5cm]{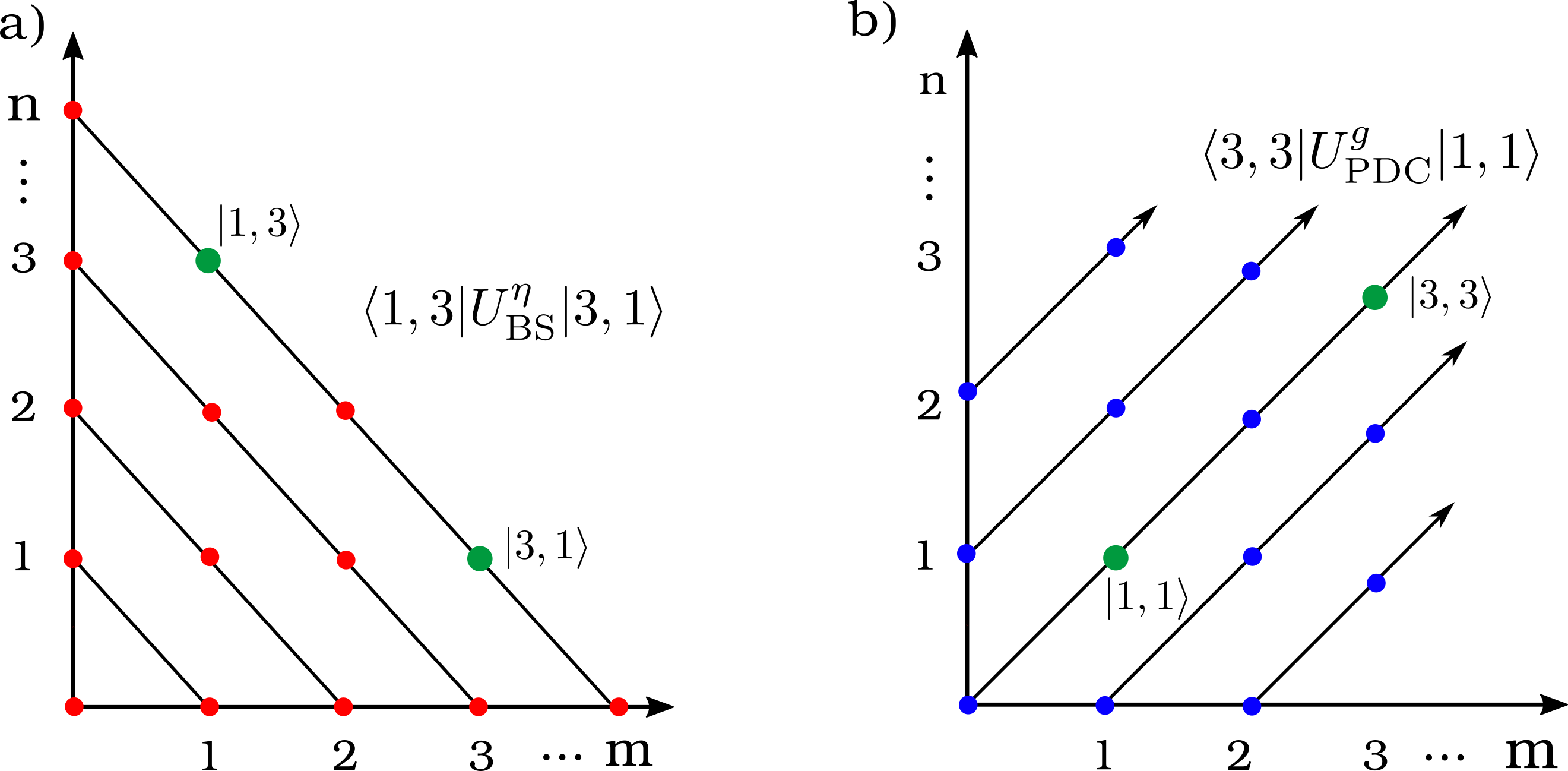}
\caption{Non-vanishing transition amplitudes between different photon-number states in  (a) a lossless beam splitter and (b) a parametric amplifier. For different numbers of photons $(n,m)$ entering each mode, one obtains different realizations of PDC and BS. For a BS, the Casimir element $J^{2}=\frac{N}{2}\left(\frac{N}{2}+1\right)$ enforces conservation of the total photon number $N=n+m$, forbidding transitions that violate this constraint. For PDC, realization of the $SU(1,1)$ Casimir, $K^{2}=\frac{n-m}{2}\left(\frac{n-m}{2}+1\right)$, instead forbids transitions that change the imbalance between the first and second modes (blue lines). The PDC/BS duality relates both representations by swapping the Casimirs; that is, it reflects  the finite-dimensional patterns in panel (a) through the vertical axis to obtain panel (b).}
\label{fig: BS_PDC_lattice}
\end{figure}

The different behavior of a parametric amplifier compared to a lossless beam splitter is illustrated in Fig.~\ref{fig: BS_PDC_lattice}. For a given initial state, a PDC produces a superposition of states that preserve the photon-number imbalance between the two modes. The number of states compatible with a fixed imbalance is unbounded, so the resulting superposition contains infinitely many terms (see the diagonal blue lines in Fig.~\ref{fig: BS_PDC_lattice}(b)). In contrast, a lossless beam splitter preserves the total photon number, and the corresponding superposition involves only a finite set of states (see Fig.~\ref{fig: BS_PDC_lattice}(a)). By introducing a parametric amplifier up to order $q$, we effectively impose a cutoff that renders the superposition finite.

A natural question concerns the accuracy of truncating the PDC expansion at order $q$. 
Since the exact vacuum amplitudes satisfy 
$c_l(g) = \tanh^{l}(\varphi)/\sqrt{g}$ with $g=\cosh^2\varphi$, 
the total weight of the discarded tail is
\begin{equation}
\varepsilon_q(g) 
= \sum_{l=q+1}^{\infty} |c_l(g)|^{2} 
= \frac{\tanh^{2(q+1)}\!\varphi}{g\,\left(1 - \tanh^{2}\varphi\right)}.
\end{equation}
For any fixed gain $g$, this quantity decreases exponentially fast with~$q$. 
Thus the truncated operation $U_{\text{PDC},q}^{g}$ reproduces the exact 
vacuum distribution with an error that can be made arbitrarily small by increasing $q$. 
In particular, for moderate gains the distribution remains strongly concentrated around $l=0$, 
so low values of $q$ already provide an excellent approximation.

Under these conditions, $U_{\text{PDC},q}^{g}$ admits a finite-dimensional matrix representation. What is not immediate, however, is that this representation can be realized as a unitary built from a universal gate set. The operator $U_{\text{PDC},q}^{g}$ is exactly unitary on the truncated subspace 
\(\mathcal{H}_{q} = \mathrm{span}\{\ket{l,l}\}_{l=0}^{q}\) (or, more generally, 
$\mathrm{span}\{\ket{n-m+l,l}\}_{l=0}^{q}$ for fixed imbalance). 
Within this subspace the map is norm preserving because the amplitudes 
$|\tilde{c}_{l}(g)|^{2}$ reproduce the exact PDC distribution restricted to 
$\{0,\dots,q\}$. 
Outside this subspace, $U_{\text{PDC},q}^{g}$ acts as the identity. 
Therefore the overall transformation is unitary when expressed as 
\[
U_{\text{PDC},q}^{g}
= U_{\text{PDC},q}^{g}\big|_{\mathcal{H}_{q}}
\;\oplus\;
1\big|_{\mathcal{H}\ominus\mathcal{H}_{q}}.
\]
This decomposition guarantees that the truncated parametric amplifier 
can be implemented as a finite-dimensional unitary, provided that one can 
coherently select the appropriate photon-number block.

To argue for unitarity, we make use of the BS-PDC duality between matrix elements in Eq.~\eqref{eqn:duality_matrix_elements}. On the multiparticle photonic Fock space, the beam splitter unitary decomposes as a direct sum of finite-dimensional blocks $U_{\text{BS},[n]}^{\eta}$, each acting on a subspace with fixed total photon number $n$:
\begin{equation}
U_{\text{BS}}^{\eta}=\bigoplus_{n=0}^\infty U_{\text{BS},[n]}^{\eta}.
\end{equation}
For a state $\ket{l,l}$ with total photon number $2l$, the BS-PDC duality in Eq.~\eqref{eqn:duality_matrix_elements} implies that the transition amplitudes of a parametric amplifier with gain $g$ can be obtained from the matrix elements of $U_{\text{BS},[2l]}^{\eta}$. Since each term in the superposition of Eq.~\eqref{eq:pdc_up_to_q} has a different total photon number, one must coherently switch among different blocks $U_{\text{BS},[2l]}^{\eta}$ depending on the input sector. Operationally, this can be achieved by introducing an ancilla that controls the relevant beam splitter representation. Once this controlled decomposition is in place, the BS-PDC duality guarantees that the truncated action in Eq.~\eqref{eq:pdc_up_to_q} can be lifted to a unitary acting on the finite-dimensional subspace defined by the cutoff $q$.

\begin{figure*}
\begin{center}
\centering
\includegraphics[width=14cm]{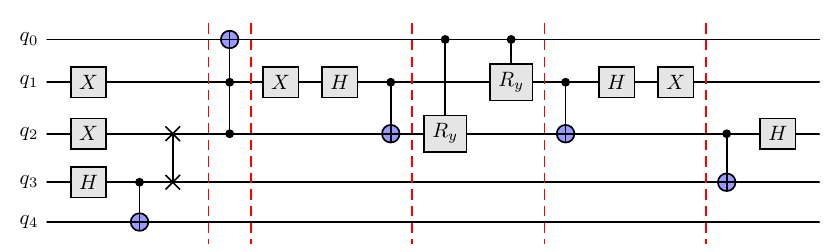}
\caption{Circuit-model implementation of the parametric amplifier with $q=1$. Barriers divide the circuit into steps corresponding to each part of the algorithm discussed in Section~\ref{sec: Diagramatic_intuition}. In the first step, an EPR pair is initialized on qubits $q_{3}$ and $q_{4}$, and the state of $q_{3}$ is swapped with that of $q_{2}$. These operations are required for the teleportation part of the protocol. The main qubits ($q_{1}$ and $q_{2}$) are then coupled to the ancilla $q_{0}$ via a CCNOT gate. The ancilla controls the corresponding BS action in the composite encoding for each fixed total number of photons entering the device. For $\ket{0,0}$ we obtain the identity, while for $\ket{1,1}$ we recover the double $R_{y}$ rotation in Eq.~\eqref{eqn: BS_ry_rotation}. Finally, a Bell-basis measurement is performed on qubits $q_2$ and $q_3$, corresponding to the caps in the diagrammatic algorithm of Fig.~\ref{fig:Circuit_scheme}.}
\label{fig:Up_to_one_PDC}
\end{center}
\end{figure*}

For the simplest non-trivial case, namely the parametric amplifier up to the generation of a single entangled photon pair $U_{\text{PDC},1}^{g}$, the circuit model implementation is shown in Fig.~\ref{fig:Up_to_one_PDC}. In this case, a total of five qubits is required: $q_{1}$ and $q_{2}$ encode the vacuum and the one-photon-per-mode state, e.g. $\ket{0,0},\ket{1,1}$; $q_{3}$ and $q_{4}$ are used to create the EPR pair required for teleportation; and $q_{0}$ is an ancilla that, as mentioned above, guarantees the unitarity of $U_{\text{PDC},q}^{g}$ by controlling the transition between the zero-particle block $U_{\text{BS},[0]}^{\eta}\cong 1$ and the two-particle block $U_{\text{BS},[2]}^{\eta} \cong R_y(\theta) \otimes R_y(\theta)$ of the BS.

\begin{table}
\begin{center}
\begin{tabular}{ |c|c| } 
 \hline
Beam splitter & Parametric amplifier \\ [0.5ex] 
 \hline 
$\braket{0,0|U_{\text{BS}}^{\eta}|0,0}=1$ & $\braket{0,0|U_{\text{PDC}}^{g}|0,0}=1/\sqrt{g}$ \\ [0.5ex]
$\braket{1,0|U_{\text{BS}}^{\eta}|0,1}=\sqrt{1-\eta}$ & $\braket{1,1|U_{\text{PDC}}^{g}|0,0}=\sqrt{g-1}/{g}$ \\ [0.5ex]
$\braket{0,1|U_{\text{BS}}^{\eta}|0,1}=\sqrt{\eta}$ & $\braket{0,1|U_{\text{PDC}}^{g}|0,1}={1}/{g}$ \\ [0.5ex] 
 $\braket{1,0|U_{\text{BS}}^{\eta}|1,0}=\sqrt{\eta}$ & $\braket{1,0|U_{\text{PDC}}^{g}|1,0}={1}/{g}$ \\ [0.5ex]
 $\braket{2,0|U_{\text{BS}}^{\eta}|1,1}=\sqrt{2\eta(1-\eta)}$ & $\braket{2,1|U_{\text{PDC}}^{g}|1,0}=\sqrt{2(g-1)}/g^{3/2}$ \\ [0.5ex]
 $\braket{0,1|U_{\text{BS}}^{\eta}|1,0}=\sqrt{1-\eta}
 $ & $\braket{0,0|U_{\text{PDC}}^{g}|1,1}=\sqrt{g-1}/{g}$ \\ [0.5ex] 
 $\braket{1,1|U_{\text{BS}}^{\eta}|1,1}=(2\eta-1)$ & $\braket{1,1|U_{\text{PDC}}^{g}|1,1}=(2-g)/g^{3/2}$ \\ [0.5ex] 
 \hline 
\end{tabular}
\caption{Exact transition amplitudes for a parametric amplifier of gain $g$ are compared to the corresponding amplitudes for a beam splitter of transmittance $\eta = g^{-1}$. The equalities between these transition amplitudes are provided by the duality 
induced by the Wick rotation between the two optical devices, as discussed in Sec.~\ref{sec: Mathematical_argument}.}
\label{table:1}
\end{center}
\end{table}

The resource requirements for implementing $U_{\text{PDC},q}^{g}$ grow in two distinct ways. First, the symmetric encoding of states with total photon number up to $2q$ requires $\mathcal{O}(q)$ physical qubits and a symmetrization stage whose ancilla cost scales quadratically with the number of encoded particles~\cite{Eckert/symmetrization/1997}. 
Second, because each term in the superposition belongs to a different photon-number sector, a coherent control register must select among the unitaries 
$U_{\text{BS},[0]}^{1/g}, U_{\text{BS},[2]}^{1/g},\dots,U_{\text{BS},[2q]}^{1/g}$. 
This requires an additional $\mathcal{O}(\log q)$ qubits and a number of controlled operations that grows linearly with $q$. Thus the overall gate complexity of the truncated PDC scales polynomially with $q$, making small values of $q$ experimentally accessible on near-term devices.

\begin{figure}[htp]
\centering
\includegraphics[width=0.95\linewidth]{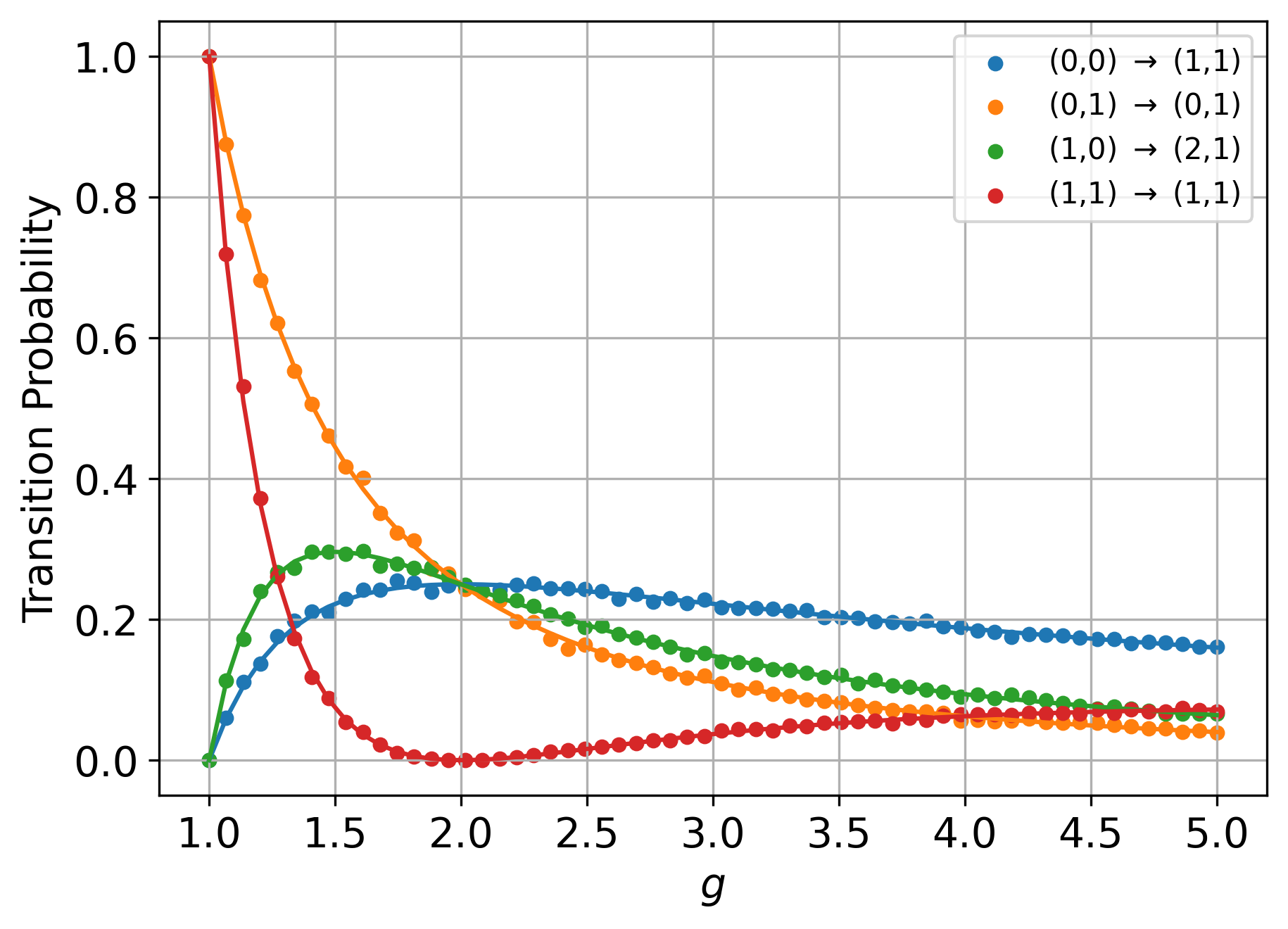}
\caption{Transition probabilities for $U_{\text{PDC},1}^{g}$ obtained from the quantum circuit in Fig.~\ref{fig:Up_to_one_PDC} and compared with the exact analytical values computed with $U_{\text{PDC}}^{g}$ and listed in Table~\ref{table:1}. Each curve displays the calculated or simulated value for different parametric gains, $g$. 
}
 \label{fig:results}
\end{figure}

By simulating the circuit-model implementation of the parametric amplifier up to $q=1$ shown in Fig.~\ref{fig:Up_to_one_PDC}, we evaluate different parametric-amplifier transition probabilities for several values of the gain $g$ using only the BS unitary. Figure~\ref{fig:results} summarizes the main results, which are in good agreement with the exact analytical values predicted by the BS-PDC duality and reported in Table~\ref{table:1}.
An interesting feature in Fig.~\ref{fig:results} is the minimum in the probability amplitude $\braket{1,1|U_{\text{PDC},1}^g|1,1}$ at $g=2$. From the BS perspective, this corresponds to the Hong-Ou-Mandel (HOM) interference dip arising from the indistinguishability of two photons entering a BS with transmittance $\eta = 1/2$~\cite{Bouchard/HOM/2021,Hong/original_HOM/1987,Jin/spectral_HOM/2015}. This result links the active nature of the parametric down-conversion dip with the passive nature of the BS-induced HOM dip. From the PDC viewpoint, the dip is characterized by a reduction in the two-photon coincidence rate due to interferometric suppression. The connection between these two phenomena highlights the central role of indistinguishability in quantum interference and entanglement, and underscores the importance of understanding the fundamental principles of quantum optics.

\section{Conclusions}\label{sec:conclusions_outlook}

In this work, we revisited the Lie group structure of beam splitters and optical parametric amplifiers and offered a geometric interpretation of their $SU(2)$ and $SU(1,1)$ descriptions. Within this framework, we demonstrated that the one-parameter families of unitaries that characterize both devices are connected by an imaginary-time (Wick) rotation on their respective group manifolds.  This correspondence identifies lossless BSs with transmittance $\eta$ as the Wick-rotated counterparts of parametric amplifiers with gain $g = 1/\eta$ and yields an exact relation between their matrix elements. This amplitude-level duality is the conceptual backbone of the circuit construction developed in this work.

We provided a diagrammatic interpretation of this duality using tensor-network diagrams and constructed an algorithm that computes PDC amplitudes by encoding photonic occupation numbers into qubit registers. 
The protocol uses a teleportation-based mechanism to exchange the encoded occupation number in one of the modes. This implements the swap required by the duality without violating causality at the physical level. 
Based on these findings, we introduced the concept of a \emph{$q$-PDC} gate, $U_{\mathrm{PDC},q}^{g}$, defined as a finite-dimensional unitary that reproduces exactly the first $q$ transition amplitudes of an ideal parametric amplifier $U_{\mathrm{PDC}}^{g}$ in the Fock basis. Since the tail of the PDC distribution exponentially decays  with photon number, the truncation error rapidly decreases with~$q$, and low values of $q$ can already provide accurate approximations for moderate gains.

We constructed and simulated the simplest nontrivial instance---the $q=1$ PDC gate---using only BS unitaries encoded within a composite map. 
The resulting transition probabilities agree with the exact analytical amplitudes predicted by the BS--PDC duality across a broad range of gains,~$g$. 

Notably, the characteristic dip at $g = 2$ naturally emerges in the circuit model of the $\langle 1,1|U_{\mathrm{PDC},1}^{g}|1,1\rangle$ channel, revealing a direct correspondence between the Hong--Ou--Mandel interference dip in lossless beam splitters and its active analogue in parametric down-conversion~\cite{Hong/original_HOM/1987,Bouchard/HOM/2021,Jin/spectral_HOM/2015}. This illustrates how indistinguishability-driven interference phenomena consistently manifest  across passive and active optical transformations.

Our results pave the way for realizing PDC-like gates on non-photonic, qubit-based architectures. This enables the implementation of truncated Gaussian-like entangling resources or boson-sampling-inspired primitives to be implemented without requiring genuine optical nonlinearities. 
The overall resource overhead scales polynomially with the truncation order $q$: increasing $q$ demands larger symmetric encodings and controlled access to the finite-dimensional BS blocks $U_{\mathrm{BS},[2l]}^{1/g}$, 
which determine the number of ancillas and the circuit depth together.  
These considerations suggest that small-$q$ implementations may be feasible in near-term platforms, while higher photon-number truncations will require further optimization of encoding and symmetrization strategies~\cite{Eckert/symmetrization/1997,Moffat/Binary_Encoding/2008}.

Beyond the single-mode-pair setting, our Wick-rotation framework raises interesting prospects for multimode generalizations. In multimode architectures, superselection rules can suppress transition amplitudes 
based on permutation symmetry at the input ports~\cite{Bezerra/HOM_superselection_rules/2023}. In the PDC picture, this  translates into structured patterns of interferometric cancellation. 
Since multimode transition amplitudes in linear optics are closely connected to matrix permanents, whose exact evaluation is NP-complete~\cite{Scheel/permanents/2004,Aaronson/permanents/2011}, it is natural to ask how the BS-PDC duality extends to full interferometric networks. Establishing a correspondence between the counting statistics of beam-splitter networks, such as the Zeilinger architecture~\cite{Zelinger/BS_network/1994}, and analogous networks of parametric amplifiers may help clarify how bosonic interference patterns, complexity-theoretic features, and entanglement-generation mechanisms transform under the Wick rotation. Exploring these connections is  a promising direction for future research.

\begin{acknowledgments}
We acknowledge partial support from the Norwegian
Ministry of Education and Research through the QTECNOS consortium (NORPART 2021-10436/CI 71331). We also acknowledge the use of open-source SDK for quantum development (Qiskit)~\cite{providers}.
\end{acknowledgments}

\bibliography{Bibliography}

\appendix

\section{Methods}

\subsection{Jordan-Schwinger representation and Wick-rotation}
\label{appendix:Lie-Group-structure}

As mentioned in the main part of the paper, two independent copies of the Bosonic algebra can be used to build realizations of the $\mathfrak{su}(2)$ and $\mathfrak{su}(1,1)$ algebras. Specifically, for the $\mathfrak{su}(2)$ case one defines the generators to be:
\begin{subequations}
\begin{align}
 J_{x} &= \frac{1}{2}(a^{\dagger}b+ab^{\dagger}),\\ J_{y} &= \frac{1}{2i}(a^{\dagger}b-ab^{\dagger}), \\
 J_{z} & = \frac{1}{2}(a^{\dagger}a-b^{\dagger}b),
\end{align}
\end{subequations}
with quadratic Casimir invariant $J^{2}=\frac{N}{2}\left(\frac{N}{2}+1\right)$. Here $N=n_{1}+n_{2}$ denotes the total number of photons entering on both modes. This particular map is due to Schwinger and has been widely used in quantum optics and in the description of two-well Bose condensates (see~\cite{Klauder/su2_interferometry/1986,Cambell/dimmer/2013}). For the case $\mathfrak{su}(1,1)$ one defines instead
\begin{subequations}
\begin{align}
 K_{x} &= \frac{1}{2}(a^{\dagger}b^{\dagger}+ab), \\
 K_{y} &= \frac{1}{2i}(a^{\dagger}b^{\dagger}-ab), \\
 K_{z} &= \frac{1}{2}(a^{\dagger}a+b^{\dagger}b+1)
\end{align}
\end{subequations}
to be the generators. For the $\mathfrak{su}(1,1)$ case, the quadratic Casimir is now $K^{2}=J_{z}(J_{z}+1)$, which is proportional to the imbalance, $n_{1}-n_{2}$, of photons entering both modes. The direct physical consequence of the difference between the Casimir elements is that while the beam splitter (described by SU(2) unitaries) preserves the total number of bosons entering the modes, parametric amplifiers (SU(1,1) unitaries) preserve the difference between the number of bosons per mode. The complexification of both algebras is isomorphic to $Sl(2,\mathbb{C})$, \emph{i.e.}
\begin{subequations}
\begin{align}
 [J_{+},J_{-}]&=2J_{z}\quad \text{and} \\ [K_{+},K_{-}]&=-2K_{z}.
\end{align}
\end{subequations}
Therefore, the representation theory is constructed similarly. For instance, in $\mathfrak{su}(1,1)$, one selects the states $\ket{j\, m}_{\mathfrak{su}(1,1)}$, where $j$ represents the eigenvalues of the Casimir $K_{z}$ and $m$ represents the eigenvalues of $K_{z}$. Whereas in $\mathfrak{su}(2)$, the usual angular momentum states $\ket{j\, m}_{\mathfrak{su}(2)}$ are chosen. However, the states $\ket{j\, m}_{\mathfrak{su}(1,1)}$ have no upper bound, resulting in an infinite-dimensional Hilbert space spanned by them.

At the level of the $\mathfrak{su}(2)/\mathfrak{su}(1,1)$ Lie-algebras, the Wick rotation is realized by making the replacements: $K_{y} \to iJ_{y}$, $K_{x} \to iJ_{x}$, and $K_{z}\to J_{z}$. As we will demonstrate shortly, this transformation results in the duality between a beam splitter and a parametric amplifier mentioned in the main content of this document. Specifically, by making use of the dissentangling formulas~\cite{DasGupta/Dissentangling/1996,Fujii/dissentangling/1999} to factorize the respective unitary of each optical device into a product of exponentials, the unitary of a parametric amplifier of gain $g=\cosh(\phi)$ can be expressed like
\begin{align}
\label{eq:PDC_dissentangling_formulae}
 U_{\text{PDC}}^{g} & = e^{\phi(a^{\dagger}b^{\dagger}-ab)} \\
 &= e^{K_{+}\tanh(\phi)}\left(\frac{1}{\cosh{\phi}}\right)^{2 K_{z}}e^{-K_{-}\tanh(\phi)}\nonumber.
\end{align}
Whereafter, following a Wick-rotation $\phi \to i \theta$, it is transformed into
\begin{equation}
    U_{\text{PDC}}^{g} \to e^{J_{+}\tan(\theta)}\left(\frac{1}{\cos{\theta}}\right)^{2 J_{z}}e^{-J_{-}\tan(\theta)}\nonumber.
\end{equation}
This result can be understood as the representation (in terms of the disentangling formula) of the unitary for a Beam-splitter of transmittance $\eta=\cos^{2}(\theta)$. It should be noted that the aforementioned Wick-rotation not only alters the $\mathfrak{su}(1,1)$ algebra generators but also transforms the Casimir element (a component of the universal enveloping algebra), that is, in relation to the two-mode Bosonic-operators,
\begin{multline}
K^{2}=\frac{a^{\dagger}a-b^{\dagger}b}{2}\left(\frac{a^{\dagger}a-b^{\dagger}b}{2}+1\right) \\ \longrightarrow \; J^{2}=\frac{a^{\dagger}a+b^{\dagger}b}{2}\left(\frac{a^{\dagger}a+b^{\dagger}b}{2}+1\right).
\end{multline}
In particular, the Wick-rotation 
at the level of the Casimir elements
indicates that the non-vanishing transition amplitudes of the beam splitter unitary  transform into the non-vanishing transition amplitudes of the parametric amplifier,
\begin{equation}
    \braket{l,m|U^{g}_{\text{PDC}}|p,q} \to \braket{l,q|U^{\eta}_{\text{BS}}|p,m}.
     \label{eq:transform}
\end{equation}
For the parametric amplifier case, the Casimir is proportional
to the photon-number difference  between modes, $l-m=p-q$. 
For the beam-splitter case, however, the total photon number, $l+q=m+p$, is the conserved quantity (see Fig.~\ref{fig: BS_PDC_lattice}).

The transformation law Eq.~\eqref{eq:transform}, which indicates a relationship between the non-vanishing transition amplitudes of a parametric amplifier and a beam-splitter after the Wick-rotation, suggests investigating a duality connecting the transition amplitudes of a PDC and a BS. Specifically, we anticipate something of the form
\begin{equation}
    \braket{l,m|U^{g}_{\text{PDC}}|p,q} = \omega(g) \braket{l,q|U^{\eta(g)}_{\text{BS}}|p,m},
\end{equation}
with $\omega(g)$ being a function that depends only on the parametric gain, and $\eta(g)$ being the dependence of the BS transmittance on $g$. We then fix $\omega(g)$ by  recalling that a BS of arbitrary transmittance leaves the two-mode vacuum state invariant, that is
\begin{equation} \omega(g)=\braket{0,0|U^{g}_{\text{PDC}}|0,0} = \frac{1}{\sqrt{g}}.
\end{equation}
However, determining the transmittance $\eta$ dependence on the parametric gain is a more involved task. Initially, we can compare the unitarity of the beam splitter with that of a parametric down-converter, and note that they both coincide in the limits of $\eta \to 1^{-}$ and $g \to 1^{+}$,
\begin{equation}
    U^{\eta}_{\text{BS}}|_{\eta \to 1^{-}}=\,U^{g}_{\text{PDC}}|_{g \to 1^{+}},
\end{equation}
where the equality sign must be interpreted under the Wick-rotation action~\footnote{This is equivalent to state that a lossless beam-splitter and a parametric amplifier of  transmittance and parametric gain one, respectively, coincide with the identity.}. Now, the transmittance ranges from $0 \leq\eta \leq1$, while the gain $g$
within the unitary representation of the parametric amplifier ranges in $1\leq g$, that is, the high-gain regime. By comparing the two contrasting regimes, namely, low transmittance with $\eta \sim 0$ and infinite parametric gain with $g\to \infty$, it is straightforward show that both regimes coincide when $\eta =\frac{1}{g}$. Particularly, the last agreements between the two previously mentioned regimes, \emph{i.e.} low transmittance ($\eta \sim 0$) with infinite parametric-gain ($g\to \infty$), and high transmittance ($\eta = 1$) with parametric gain ($g=1$) via the Wick-rotation and the identification $\eta=\frac{1}{g}$ allows us to state, 
\begin{equation}
    \braket{l,m|U^{g}_{\text{PDC}}|p,q} = \frac{1}{\sqrt{g}} \braket{l,q|U^{\frac{1}{g}}_{\text{BS}}|p,m}.
\end{equation}

\subsection{Beam splitter and parametric amplifier matrix elements}

For completeness, this appendix calculates the exact matrix elements shown in Table~\ref{table:1}, for a parametric amplifier with gain $g$ and a beam splitter with transmittance $\eta$. This calculation utilizes the disentangling formulas from Section~\ref{appendix:Lie-Group-structure}, which have already been expressed as a product of normally ordered exponentials.  Specifically, for a parametric amplifier acting over the vacuum, the disentangling-formula~\eqref{eq:PDC_dissentangling_formulae} gives 
\begin{equation}
\braket{0,0|U_{\text{PDC}}^{g}|0,0}=\frac{1}{\cosh{(\phi)}}=\frac{1}{\sqrt{g}}.
\end{equation}
More generally, the transition amplitudes from the vacuum to upper states of $q$-pairs of entangled photons, can also be exactly computed. These transition amplitudes were utilized in defining the parametric amplifier up to $q$, which was discussed in section \ref{sec: pdc_concept}. We obtain:
\begin{align}
U_{\text{PDC}}^{g}\ket{0,0} &= \frac{1}{\cosh{(\phi)}}\sum_{l=0}\frac{\tanh^{l}(\phi)}{l!}a^{\dagger \, l}b^{\dagger \,l}\ket{0,0} \nonumber \\
 & =\frac{1}{\cosh{(\phi)}}\sum_{l=0}\tanh^{l}(\phi)\ket{l,l}.
\end{align}
For initial states other than  the vacuum, applying the rightmost exponential within Eq.~\eqref{eq:PDC_dissentangling_formulae} results in a finite superposition of states that have a decreasing number of photons.
For instance, in the case of the state $\ket{1,1}$, we have $\exp(-ab \tanh{\phi})\ket{1,1}=\ket{1,1}-\tanh(\phi)\ket{0,0}$, and the transition probability reads
\begin{equation}
\braket{1,1|U_{\text{PDC}}^{g}|1,1}=\frac{1}{\cosh^{3}(\phi)}-\frac{\tanh^{2}(\phi)}{{\cosh^{3}(\phi)}}=\frac{2-g}{g^{\frac{3}{2}}},
\end{equation}
whereas for the $\ket{0,1}$ state we can
similarly show that 
\begin{equation}
\braket{0,1|U_{\text{PDC}}^{g}|0,1}=\frac{1}{\cosh^{2}(\phi)}=\frac{1}{g}.
\end{equation}
The disentangling formulae for a beam splitter with transmittance $\eta=\cos^{2}(\theta)$ leads to 
\begin{equation}
\label{eq:BS-dissentangling}
    U_{\text{BS}}^{\eta}= e^{J_{+}\tan(\theta)}\left(\frac{1}{\cos{\theta}}\right)^{2 K_{z}}e^{-J_{-}\tan(\theta)}\nonumber.
\end{equation}
As the two-mode vacuum is invariant under the action of a beam-splitter, 
\begin{equation}
    \braket{0,0|U_{\text{BS}}^{\eta}|0,0}=1,
\end{equation}
whereas for states different from the vacuum, the action of the beam-splitter creates superposition of states with equal number of photons:
\begin{equation}
    U_{\text{BS}}^{\eta}\ket{m,n}=\sum_{l=0}^{m}c_{l}\ket{m-l,n+l},
\end{equation}
and $n\leq m$. Particularly, for the $\ket{1,1}$ and $\ket{0,1}$ states, the beam-splitter disentangling formulae \eqref{eq:BS-dissentangling} gives
\begin{equation}  \braket{1,1|U_{\text{BS}}^{\eta}|1,1}=(2\eta -1)\quad \text{and}\quad \braket{0,1|U_{\text{BS}}^{\eta}|0,1}=0,
\end{equation}
where for the 50:50 beam-splitter case, $\eta=\frac{1}{2}$, we retrieve the celebrated Hong-Ou-Mandel dip associated with the cancellation of the $\ket{1,1}$ transition amplitude due to photon indistinguishably.

\subsection{Optical Observables}

\subsubsection{Mean number of photons for a q-parametric amplifier}

\begin{figure}
\includegraphics[width=0.9\linewidth]{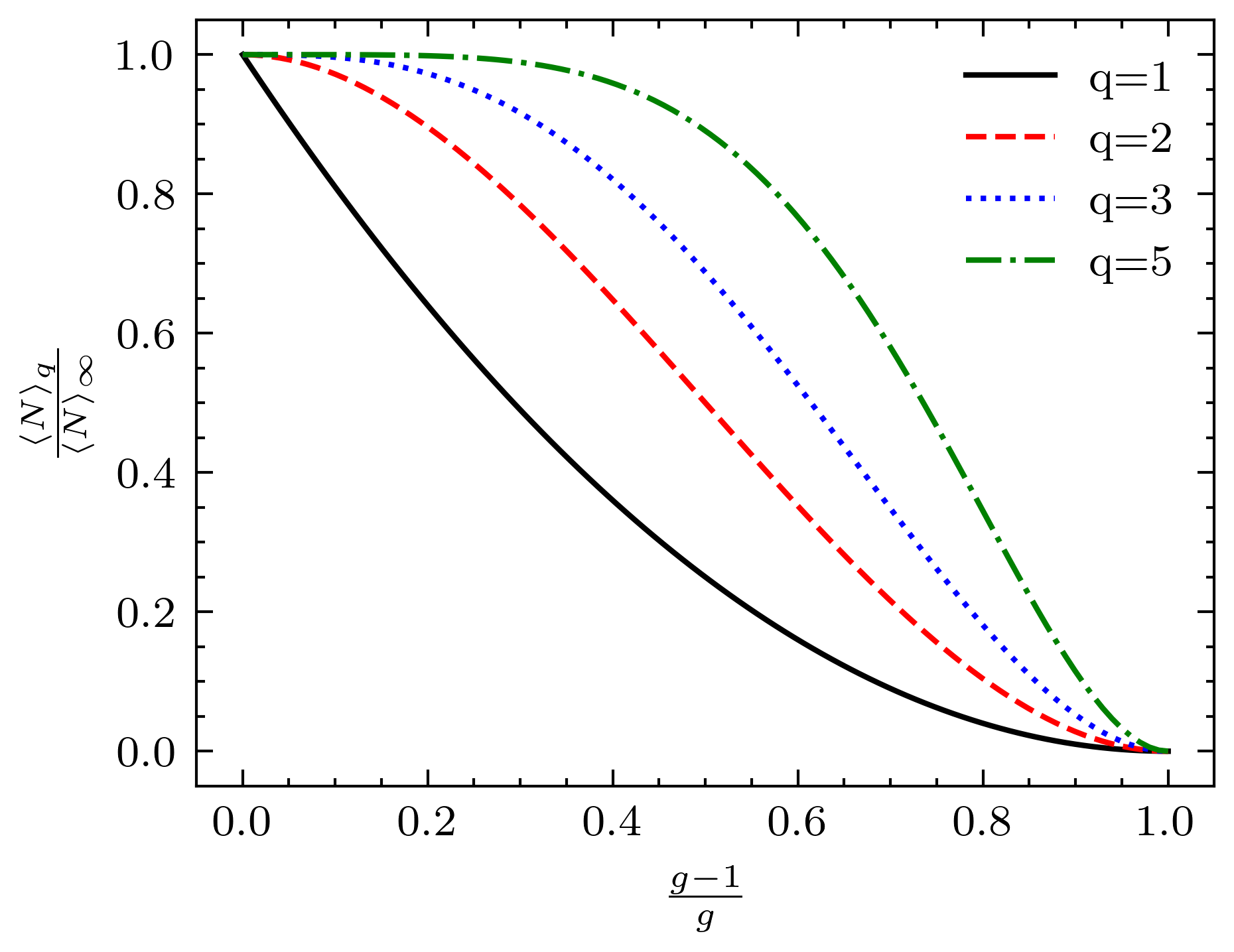}
    \caption{Ratio of the mean number of photons produced by the parametric amplifier up to $q$ to the actual mean number of photons, within the high gain regime, $g\geq 1$.}
\label{fig:Mean_number_photons}
\end{figure}

We evaluate the mean photon occupation number $\braket{N}_{q}$ for a parametric amplifier up to the creation of $q$-entangled photon pairs:
\begin{equation}
 \braket{N}_{q} = \braket{0,0|U_{\text{PDC},q}^{g \dagger}\,N\,U_{\text{PDC},q}^{g}|0,0}.
\end{equation}
In the limit $q \to \infty$, we obtain the genuine mean-photon occupation number:
\begin{equation}
 \braket{N}_{\infty} = \braket{0,0|U_{\text{PDC}}^{g \dagger}N\,U_{\text{PDC}}^{g}|0,0}=2(g-1).
\end{equation}
As the parametric amplifier only allows transition amplitudes between states conserving the imbalance between both modes, the expected value is given by
\begin{multline}
\label{eq:mean_number_of_photons}
 \braket{N}_{q} = \sum_{l=0}^{q}2l |\braket{l,l|U_{\text{PDC},q}^{g}|0,0}|^{2} = \sum_{l=0}^{q} 2l \frac{\tanh{(\phi)}^{2l}}{\cosh{(\phi)}^{2}}.
\end{multline}
As stated in the main text, the parametric gain $g$ relates to the angle $\phi$ as $g=\cosh{(\phi)}^{2}$. Furthermore, the final sum can be exactly computed by setting it in terms of a geometric sum:
\begin{multline}
 \sum_{l=0}^{q} 2l \frac{\tanh{(\phi)}^{2l}}{\cosh{(\phi)}^{2}}=\tanh{\phi} \frac{\mathrm{d}}{\mathrm{d}\phi} \left(\sum_{l=0}^{q} \tanh{(\phi)}^{2l} \right) \\ =\tanh{\phi} \frac{\mathrm{d}}{\mathrm{d}\phi} \left( \frac{1-\tanh{(\phi)^{2(q+1)}}}{1-\tanh{(\phi)^{2}}}\right) .
\end{multline}
From this result, the 
ratio of the mean number of photons generated by a parametric amplifier up to $q$ and the mean number of photons generated by the complete action of the parametric amplifier yields
\begin{align}
 \frac{\braket{N}_{q}}{\braket{N}_{\infty}} & = 1 - (1+q)\left(\tanh{^{2}(\phi)}\right)^{q} + q\left(\tanh{^{2}(\phi)}\right)^{q+1} \notag \\ 
 & = 1 - (1+q)\left(\frac{g-1}{g}\right)^{q} + q\left(\frac{g-1}{g}\right)^{q+1}.
\end{align}
Figure~\ref{fig:Mean_number_photons} displays this ratio within the high gain regime ($g\geq 1$). For $g\to \infty$ (i.e., $\frac{g-1}{g}\to 1$), the expected mean number of photons cannot be produced by the parametric amplifier unless $q\to \infty$. Nevertheless, with moderately high gain ($g=2$), using a parametric amplifier up to $q=2$, we can recover approximately 60\% of the total mean number of photons.
\begin{figure}
\includegraphics[width=\linewidth]{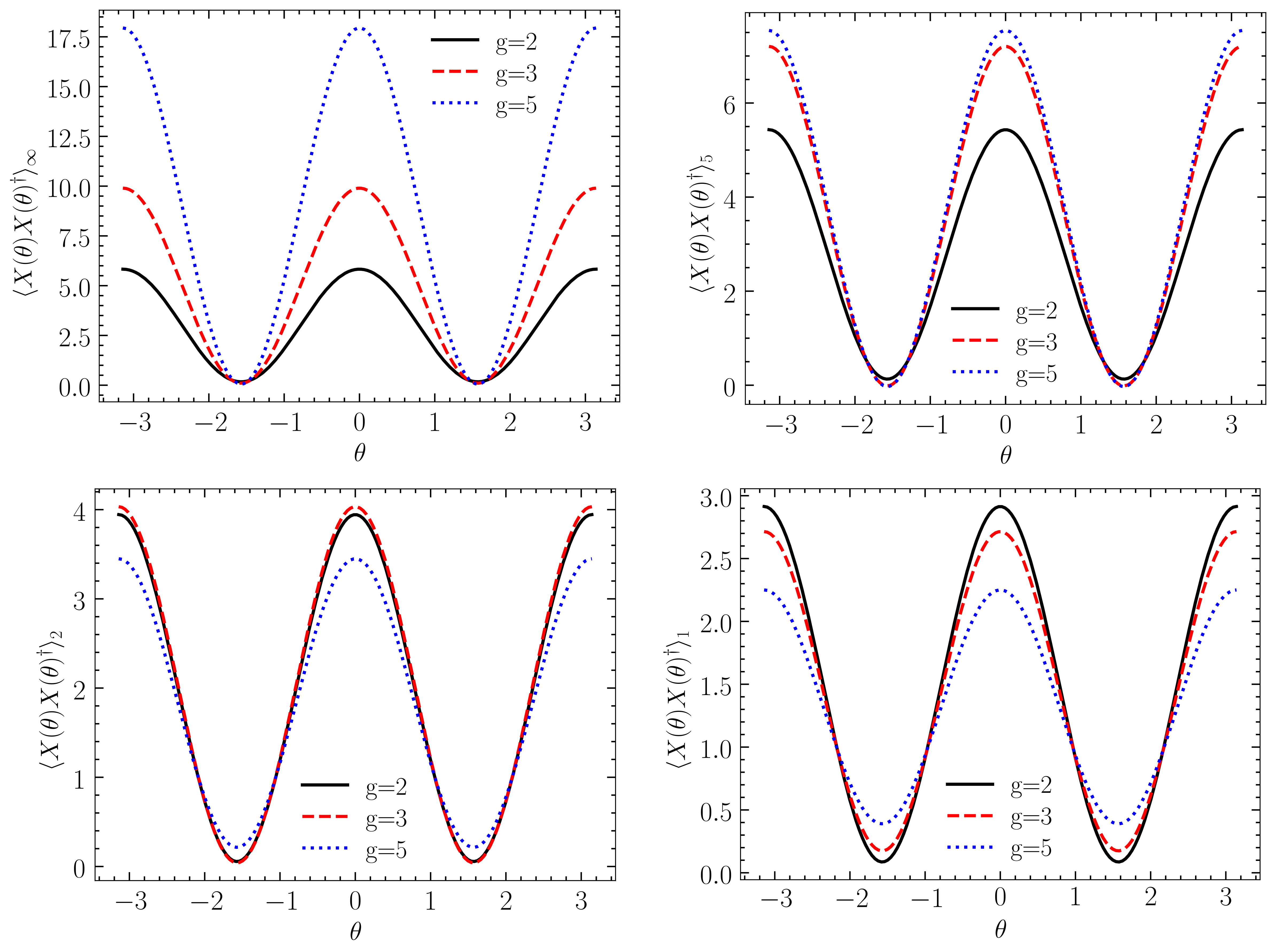}
\caption{Fluctuations in the two-mode quadrature operator $X(\theta)$ were computed for varying values of both $q$ and the parametric gain $g$. As the vacuum is used to evaluate fluctuations, $\braket{X(\theta)X(\theta)^{\dagger}}=\braket{Y(\theta)Y(\theta)^{\dagger}}$.}
    \label{fig:Homodyne_measurements}
\end{figure}

\subsubsection{Two-mode quadrature}

Let $X(\theta)$ and $Y(\theta)$ be the two-mode quadrature operators that satisfy the canonical commutation relations $[X(\theta),Y(\theta)]=2i$,
\begin{subequations}
\begin{align}
 X(\theta) & =e^{-i\theta}a+e^{i \theta}b^{\dagger}, \\ Y(\theta) &=ie^{i\theta}a^{\dagger}+ie^{-i \theta}b.
 \end{align}
\end{subequations}
The relation between the two-mode quadrature and the single-mode quadrature, $X_{a}(\theta)=e^{-i\theta}a+e^{i \theta}a^{\dagger}$ and $Y_{a}(\theta)=ie^{i\theta}a^{\dagger}+ie^{-i \theta}a$, is provided by the constraints~\cite{Shaked/two_mode_quadrature/2018}
\begin{subequations}
  \begin{align}
X(\theta)+X(\theta)^{\dagger} &=X_{a}(\theta)+X_{b}(\theta), \\ Y(\theta)+Y(\theta)^{\dagger} &=-i(Y_{a}(\theta)-Y_{b}(\theta)).
  \end{align}  
\end{subequations}
The fluctuations in the two-mode quadrature operator over a parametrically amplified vacuum up to $q$ read
\begin{align} 
\braket{X(\theta)X(\theta)^{\dagger}}_{q}&=\braket{0,0|U_{\text{PDC},q}^{g \dagger}X(\theta)X(\theta)^{\dagger}U_{\text{PDC},q}^{g}|0,0} \notag\\ &=1+\braket{N}_{q}+e^{2i\theta}\braket{a^{\dagger}b^{\dagger}}_{q}+e^{-i2\theta}\braket{ab}_{q}.
\end{align}
The fluctuations grow linearly with the mean number of photons. For the vacuum,  $\braket{a^{\dagger}b^{\dagger}}_{q}$ can be exactly evaluated as follows:
\begin{widetext}
\begin{eqnarray}
\nonumber
\braket{a^{\dagger}b^{\dagger}}_{q} & = & \sum_{l,m=0}^{q} \braket{l,l|U_{\text{PDC},q}^{g}|0,0}\braket{0,0|U_{\text{PDC},q}^{g}|m,m}\braket{m,m|a^{\dagger}b^{\dagger}|l,l}
 = \sum_{l=0}^{q} (l+1) \frac{\tanh^{2l+1}{(\phi)}}{\cosh^{2}{(\phi)}} \\
  & = & \frac{\tanh{(\phi)}}{\cosh^{2}{(\phi)}}\sum_{l=0}^{q}\left(\tanh^{2}{(\phi)}\right)^{l}\, + \,\frac{\tanh{(\phi)}}{2}\sum_{l=0}^{q} (2l) \frac{\tanh^{2l}{(\phi)}}{\cosh^{2}{(\phi)}}.
\label{eq:mean_number_abq}
\end{eqnarray}
\end{widetext}
%
%
In Eq.~\eqref{eq:mean_number_abq}, the first term is a geometric sum, whereas the last sum is proportional to the total mean number of photons Eq.~\eqref{eq:mean_number_of_photons}. By substituting  these terms, the fluctuations of the two-mode quadrature are exactly given by
\begin{widetext}
\begin{equation}
\label{eq:two_mode_fluctuations}
\braket{X(\theta)X(\theta)^{\dagger}}_{q}=1+\braket{N}_{q}+2\cos{(\theta)}\left(\frac{1}{g}\sqrt{\frac{g-1}{g}}\left(\frac{1-\left(\frac{g-1}{g}\right)^{q+1}}{1-\left(\frac{g-1}{g}\right)}\right)+\frac{1}{2}\sqrt{\frac{g-1}{d}}\braket{N}_{q}\right).
\end{equation} 
\end{widetext}
In Fig.~\ref{fig:Homodyne_measurements} we plot the profile fluctuations of the two-mode quadrature for various $q$ values. 
The left upper panel displays the fluctuations calculated using the full PDC unitary $U_{\text{PDC}}^{g}$.
The increase in parametric gain heightens the amplitude of the fluctuations as predicted by the mean-photon number dependence in Eq.~\eqref{eq:two_mode_fluctuations}. 
\end{document}

%% file: Tensors.tikzstyles
\tikzstyle{red_ball}=[fill=red, draw=red, shape=circle, tikzit category=Ball]
\tikzstyle{blue_ball}=[fill=blue, draw=blue, shape=circle, tikzit category=Ball, radius=0.2]
\tikzstyle{medium_box}=[fill=white, draw=black, shape=rectangle, tikzit category=boxes, minimum width=0.75cm, minimum height=1cm]
\tikzstyle{blank_edge}=[line width=0.2cm]
\tikzstyle{small_box}=[fill=white, draw=black, shape=rectangle, tikzit category=boxes, minimum height=0.4cm, minimum width=0.4cm]
\tikzstyle{big_box}=[fill=white, draw=black, shape=rectangle, tikzit category=boxes, minimum height=2.0cm, minimum width=1.5cm]
\tikzstyle{small_box_2}=[fill=white, draw=black, shape=rectangle, tikzit category=boxes, minimum height=0.6cm, minimum width=0.6cm]

\tikzstyle{white_edge}=[-, draw=white, line width=3.2pt]
\tikzstyle{red_arrow}=[->, draw=red]
\tikzstyle{blue_arrow}=[->, draw=blue]
\tikzstyle{arrow}=[->]
\tikzstyle{blue_filled_line}=[-, fill={rgb,255: red,255; green,193; blue,207}, draw=black]
\tikzstyle{pink_filled}=[-, fill={rgb,255: red,165; green,255; blue,235}]
\tikzstyle{blue_edge}=[-, draw=blue]
\tikzstyle{red_edge}=[-, draw=red]

%% file: Bibliography.bib
@article{Low/Chuang/Qubitization/2019,
  author    = {Low, Guang Hao and Chuang, Isaac L.},
  title     = {Hamiltonian Simulation by Qubitization},
  journal   = {Quantum},
  volume    = {3},
  pages     = {163},
  year      = {2019},
  doi       = {10.22331/q-2019-07-12-163}
}

@article{Low/QSP/2017,
  author    = {Low, Guang Hao and Chuang, Isaac L.},
  title     = {Optimal Hamiltonian Simulation by Quantum Signal Processing},
  journal   = {Phys. Rev. Lett.},
  volume    = {118},
  number    = {1},
  pages     = {010501},
  year      = {2017},
  doi       = {10.1103/PhysRevLett.118.010501}
}

@article{Su/Product_formulas/2019,
  title = {Nearly Optimal Lattice Simulation by Product Formulas},
  author = {Childs, Andrew M. and Su, Yuan},
  journal = {Phys. Rev. Lett.},
  volume = {123},
  issue = {5},
  pages = {050503},
  numpages = {6},
  year = {2019},
  month = {Aug},
  publisher = {American Physical Society},
  doi = {10.1103/PhysRevLett.123.050503},
  url = {https://link.aps.org/doi/10.1103/PhysRevLett.123.050503}
}

@article{Lloyd/universal_simulators/1996,
  author    = {Lloyd, Seth},
  title     = {Universal quantum simulators},
  journal   = {Science},
  volume    = {273},
  number    = {5278},
  pages     = {1073--1078},
  year      = {1996},
  doi       = {10.1126/science.273.5278.1073}
}

@article{Childs/Trotter_error/2021,
  author    = {Childs, Andrew M. and Su, Yuan and Tran, Minh C. and Wiebe, Nathan and Zhu, Shuchen},
  title     = {Theory of Trotter Error},
  journal   = {PRX Quantum},
  volume    = {2},
  number    = {1},
  pages     = {010329},
  year      = {2021},
  doi       = {10.1103/PRXQuantum.2.010329}
}

@misc{Preskill/quantum_advantage/2012,
Author = {J. Preskill},
Title = {Quantum computing and the entanglement frontier},
Year = {2012},
Eprint = {arXiv:1203.5813}
}

@article{Preskill/Nisq/2018,
  doi = {10.22331/q-2018-08-06-79},
  url = {https://doi.org/10.22331/q-2018-08-06-79},
  title = {Quantum {C}omputing in the {NISQ} era and beyond},
  author = {J. Preskill},
  journal = {Quantum},
  issn = {2521-327X},
  volume = {2},
  pages = {79},
  year = {2018}
}

@article{Harrow/Computational_supremacy/2017,
  doi = {10.1038/nature23458},
  url = {https://doi.org/10.1038/nature23458},
  year = {2017},
  month = sep,
  publisher = {Springer Science and Business Media {LLC}},
  volume = {549},
  number = {7671},
  pages = {203--209},
  author = {A. W. Harrow and A. Montanaro},
  title = {Quantum computational supremacy},
  journal = {Nature}
}

@article{Eisert/quantum_random_sampling_advantage/2023,
  title = {Computational advantage of quantum random sampling},
  author = {D. Hangleiter and J. Eisert},
  journal = {Rev. Mod. Phys.},
  volume = {95},
  issue = {3},
  pages = {035001},
  numpages = {82},
  year = {2023},
  month = {Jul},
  publisher = {American Physical Society},
  doi = {10.1103/RevModPhys.95.035001},
  url = {https://link.aps.org/doi/10.1103/RevModPhys.95.035001}
}

@article{Wang/Boson_samping_advantage/2023,
  title = {Experimental Boson Sampling Enabling Cryptographic One-Way Function},
  author = {Wang, Xiao-Wei and Zhou, Wen-Hao and Fu, Yu-Xuan and Gao, Jun and Lu, Yong-Heng and Chang, Yi-Jun and Qiao, Lu-Feng and Ren, Ruo-Jing and Jiang, Ze-Kun and Jiao, Zhi-Qiang and Nikolopoulos, Georgios M. and Jin, Xian-Mi},
  journal = {Phys. Rev. Lett.},
  volume = {130},
  issue = {6},
  pages = {060802},
  numpages = {6},
  year = {2023},
  month = {Feb},
  publisher = {American Physical Society},
  doi = {10.1103/PhysRevLett.130.060802},
  url = {https://link.aps.org/doi/10.1103/PhysRevLett.130.060802}
}

@Article{Biamonte/Q_machine_learnign/2017,
author={J. Biamonte and P. Wittek and N. Pancotti P. Rebentrost and N. Wiebe and S. Lloyd},
title={Quantum machine learning},
journal={Nature},
year={2017},
month={Sep},
day={01},
volume={549},
number={7671},
pages={195-202},
issn={1476-4687},
doi={10.1038/nature23474},
url={https://doi.org/10.1038/nature23474}
}

@article{pffaf/Quantum_teleportation/2018,
author = {S. Olmschenk and D. N. Matsukevich  and P. Maunz  and D. Hayes and L.-M. Duan and C. Monroe },
title = {Quantum Teleportation Between Distant Matter Qubits},
journal = {Science},
volume = {323},
number = {5913},
pages = {486-489},
year = {2009},
doi = {10.1126/science.1167209},
URL = {https://www.science.org/doi/abs/10.1126/science.1167209},
}

@article{Giovanneti/Quantum_measurents/2004,
author = {V. Giovannetti and S. Lloyd and L. Maccone},
title = {Quantum-Enhanced Measurements: Beating the Standard Quantum Limit},
journal = {Science},
volume = {306},
number = {5700},
pages = {1330-1336},
year = {2004},
doi = {10.1126/science.1104149},
URL = {https://www.science.org/doi/abs/10.1126/science.1104149},
}

@article{Omar/QSensing23,
  title = {Quantum enhanced probing of multilayered samples},
  author = {Li-Gomez, M. Y. and Yepiz-Graciano, P. and Hrushevskyi, T. and Calder\'on-Losada, O. and Saglamyurek, E. and Lopez-Mago, D. and Salari, V. and Ngo, T. and U'Ren, A. B. and Barzanjeh, Shabir},
  journal = {Phys. Rev. Res.},
  volume = {5},
  issue = {2},
  pages = {023170},
  numpages = {12},
  year = {2023},
  month = {Jun},
  publisher = {American Physical Society},
  doi = {10.1103/PhysRevResearch.5.023170},
  url = {https://link.aps.org/doi/10.1103/PhysRevResearch.5.023170}
}

@article{JH/Qadvantage17,
	Author = {Melo-Luna, C. A. and Susa, C. E. and Ducuara, A. F. and Barreiro, A. and Reina, J. H.},
	Da = {2017/03/22},
	Doi = {10.1038/srep44730},
	Id = {Melo-Luna2017},
	Isbn = {2045-2322},
	Journal = {Scientific Reports},
	Number = {1},
	Pages = {44730},
	Title = {Quantum Locality in Game Strategy},
	Ty = {JOUR},
	Url = {https://doi.org/10.1038/srep44730},
	Volume = {7},
	Year = {2017}}

@article{JH/Q_Engines23,
  title = {Correlation-boosted quantum engine: A proof-of-principle demonstration},
  author = {Herrera, Marcela and Reina, John H. and D'Amico, Irene and Serra, Roberto M.},
  journal = {Phys. Rev. Res.},
  volume = {5},
  issue = {4},
  pages = {043104},
  numpages = {11},
  year = {2023},
  month = {Nov},
  publisher = {American Physical Society},
  doi = {10.1103/PhysRevResearch.5.043104},
  url = {https://link.aps.org/doi/10.1103/PhysRevResearch.5.043104}
}

@article{Flamini/photonic_review/2019,
doi = {10.1088/1361-6633/aad5b2},
url = {https://dx.doi.org/10.1088/1361-6633/aad5b2},
year = {2018},
month = {nov},
publisher = {IOP Publishing},
volume = {82},
number = {1},
pages = {016001},
author = {F. Flamini and N. Spagnolo and F. Sciarrino},
title = {Photonic quantum information processing: a review},
journal = {Reports on Progress in Physics},
}

@Article{PhotonQP22,
author={L. S. Madsen
and F. Laudenbach
and M. F. Askarani
and others},
title={Quantum computational advantage with a programmable photonic processor},
journal={Nature},
year={2022},
month={Jun},
day={01},
volume={606},
number={7912},
pages={75-81},
issn={1476-4687},
doi={10.1038/s41586-022-04725-x},
url={https://doi.org/10.1038/s41586-022-04725-x}
}

@article{Zhong/computational_advantage_photons/2019,
author = {H. S. Zhong and H. Wang and Y. H. Deng and others},
title = {Quantum computational advantage using photons},
journal = {Science},
volume = {370},
number = {6523},
pages = {1460-1463},
year = {2020},
doi = {10.1126/science.abe8770},
URL = {https://www.science.org/doi/abs/10.1126/science.abe8770},
}

@article{Jacques/universal_optics/2019,
author = {J. Carolan and C. Harrold and C. Sparrow and others},
title = {Universal linear optics},
journal = {Science},
volume = {349},
number = {6249},
pages = {711-716},
year = {2015},
doi = {10.1126/science.aab3642},
URL = {https://www.science.org/doi/abs/10.1126/science.aab3642},
}

@article{Neil/Super_conducting_supremacy/2018,
author = {C. Neill and P. Roushan and K. Kechedzhi and others},
title = {A blueprint for demonstrating quantum supremacy with superconducting qubits},
journal = {Science},
volume = {360},
number = {6385},
pages = {195-199},
year = {2018},
doi = {10.1126/science.aao4309},
URL = {https://www.science.org/doi/abs/10.1126/science.aao4309},
}

@article{Chatterjee/super_conductin_practice/2021,
  doi = {10.1038/s42254-021-00283-9},
  url = {https://doi.org/10.1038/s42254-021-00283-9},
  year = {2021},
  month = feb,
  publisher = {Springer Science and Business Media {LLC}},
  volume = {3},
  number = {3},
  pages = {157--177},
  author = {A. Chatterjee and P. Stevenson and S. De Franceschi and A. Morello and N. P. de Leon and F. Kuemmeth},
  title = {Semiconductor qubits in practice},
  journal = {Nature Reviews Physics}
}

@article{Georgescu/ion_trap_rev/2020,
  doi = {10.1038/s42254-020-0189-1},
  url = {https://doi.org/10.1038/s42254-020-0189-1},
  year = {2020},
  month = may,
  publisher = {Springer Science and Business Media {LLC}},
  volume = {2},
  number = {6},
  pages = {278--278},
  author = {I. Georgescu},
  title = {Trapped ion quantum computing turns 25},
  journal = {Nature Reviews Physics}
}

@article{Auger/rydberg_blueprint/2017,
  title = {Blueprint for fault-tolerant quantum computation with Rydberg atoms},
  author = {J. M. Auger and S. Bergamini and D. E. Browne},
  journal = {Phys. Rev. A},
  volume = {96},
  issue = {5},
  pages = {052320},
  numpages = {6},
  year = {2017},
  month = {Nov},
  publisher = {American Physical Society},
  doi = {10.1103/PhysRevA.96.052320},
  url = {https://link.aps.org/doi/10.1103/PhysRevA.96.052320}
}

@article{JH/QDs03,
  title = {Optical schemes for quantum computation in quantum dot molecules},
  author = {Lovett, B. W. and Reina, J. H. and Nazir, A. and Briggs, G. A. D.},
  journal = {Phys. Rev. B},
  volume = {68},
  issue = {20},
  pages = {205319},
  numpages = {18},
  year = {2003},
  month = {Nov},
  publisher = {American Physical Society},
  doi = {10.1103/PhysRevB.68.205319},
  url = {https://link.aps.org/doi/10.1103/PhysRevB.68.205319}
}

@article{JH/spins04,
  title = {Optical quantum computation with perpetually coupled spins},
  author = {Benjamin, S. C. and Lovett, B. W. and Reina, J. H.},
  journal = {Phys. Rev. A},
  volume = {70},
  issue = {6},
  pages = {060305},
  numpages = {4},
  year = {2004},
  month = {Dec},
  publisher = {American Physical Society},
  doi = {10.1103/PhysRevA.70.060305},
  url = {https://link.aps.org/doi/10.1103/PhysRevA.70.060305}
}

@article{Henriet/neutral_atom_QC/2020,
  doi = {10.22331/q-2020-09-21-327},
  url = {https://doi.org/10.22331/q-2020-09-21-327},
  title = {Quantum computing with neutral atoms},
  author = {L. Henriet and L. Beguin and A. Signoles and T. Lahaye and A. Browaeys and G. O. Reymond and C. Jurczak},
  journal = {{Quantum}},
  issn = {2521-327X},
  publisher = {{Verein zur F{\"{o}}rderung des Open Access Publizierens in den Quantenwissenschaften}},
  volume = {4},
  pages = {327},
  month = sep,
  year = {2020}
}

@article{Gonzalez/silicon_based_review/2021,
  doi = {10.1038/s41928-021-00681-y},
  url = {https://doi.org/10.1038/s41928-021-00681-y},
  year = {2021},
  month = dec,
  publisher = {Springer Science and Business Media {LLC}},
  volume = {4},
  number = {12},
  pages = {872--884},
  author = {M. F. Gonzalez-Zalba and S. de Franceschi and E. Charbon and T. Meunier and M. Vinet and A. S. Dzurak},
  title = {Scaling silicon-based quantum computing using {CMOS} technology},
  journal = {Nature Electronics}
}

@Article{KML/Quantum_computing/2001,
author={E. Knill
and R. Laflamme
and G. J. Milburn},
title={A scheme for efficient quantum computation with linear optics},
journal={Nature},
year={2001},
month={Jan},
day={01},
volume={409},
number={6816},
pages={46-52},
issn={1476-4687},
doi={10.1038/35051009},
url={https://doi.org/10.1038/35051009}
}

@article{Milburn/Review/2007,
  title = {Linear optical quantum computing with photonic qubits},
  author = {P. Kok and W. J. Munro and K. Nemoto and T. C. Ralph and J. P. Dowling and G. J. Milburn},
  journal = {Rev. Mod. Phys.},
  volume = {79},
  issue = {1},
  pages = {135--174},
  numpages = {0},
  year = {2007},
  month = {Jan},
  publisher = {American Physical Society},
  doi = {10.1103/RevModPhys.79.135},
  url = {https://link.aps.org/doi/10.1103/RevModPhys.79.135}
}

@article{Klauder/su2_interferometry/1986,
  title = {SU(2) and SU(1,1) interferometers},
  author = {B. Yurke and S. L. McCall and J. R. Klauder},
  journal = {Phys. Rev. A},
  volume = {33},
  issue = {6},
  pages = {4033--4054},
  numpages = {0},
  year = {1986},
  month = {Jun},
  publisher = {American Physical Society},
  doi = {10.1103/PhysRevA.33.4033},
  url = {https://link.aps.org/doi/10.1103/PhysRevA.33.4033}
}

@article{DasGupta/Dissentangling/1996,
  doi = {10.1119/1.18183},
  url = {https://doi.org/10.1119/1.18183},
  year = {1996},
  month = nov,
  publisher = {American Association of Physics Teachers ({AAPT})},
  volume = {64},
  number = {11},
  pages = {1422--1427},
  author = {A. DasGupta},
  title = {Disentanglement formulas: An alternative derivation and some applications to squeezed coherent states},
  journal = {American Journal of Physics}
}

@Article{Ruan/Two_boson_Higgs_Algebra/2003,
author={D. Ruan},
title={Two-boson Realizations of the Polynomial Angular Momentum Algebra and Some Applications},
journal={Journal of Mathematical Chemistry},
year={2006},
month={Feb},
day={01},
volume={39},
number={2},
pages={417-440},
issn={1572-8897},
doi={10.1007/s10910-005-9025-1},
url={https://doi.org/10.1007/s10910-005-9025-1}
}

@book{sakurai/Modern/2003,
      author        = "J. J. Sakurai",
      title         = "{Modern Quantum Mechanics; rev. ed.}",
      publisher     = "Addison-Wesley",
      address       = "Reading, MA",
      year          = "1994",
      url           = "https://cds.cern.ch/record/1167961",
}

@article{Ou/Li/2020,
    author = {Z. Y. Ou and X. Li},
    title = "{Quantum SU(1,1) interferometers: Basic principles and applications}",
    journal = {APL Photonics},
    volume = {5},
    number = {8},
    year = {2020},
    month = {08},
    issn = {2378-0967},
    doi = {10.1063/5.0004873},
    url = {https://doi.org/10.1063/5.0004873},
    note = {080902},
}

@article{Jing/2011,
    author = {J. Jing and C. Liu and Z. Zhou and Z. Y. Ou and W. Zhang},
    title = "{Realization of a nonlinear interferometer with parametric amplifiers}",
    journal = {Applied Physics Letters},
    volume = {99},
    number = {1},
    year = {2011},
    month = {07},
    issn = {0003-6951},
    doi = {10.1063/1.3606549},
    url = {https://doi.org/10.1063/1.3606549},
    note = {011110},
}

@article{Chekhova:16,
author = {M. V. Chekhova and Z. Y. Ou},
journal = {Adv. Opt. Photon.},
number = {1},
pages = {104--155},
publisher = {Optica Publishing Group},
title = {Nonlinear interferometers in quantum optics},
volume = {8},
month = {Mar},
year = {2016},
url = {https://opg.optica.org/aop/abstract.cfm?URI=aop-8-1-104},
doi = {10.1364/AOP.8.000104},
}

@article{Cerf/Two_boson_interference/2019,
author = {N. J. Cerf and M. G. Jabbour },
title = {Two-boson quantum interference in time},
journal = {Proceedings of the National Academy of Sciences},
volume = {117},
number = {52},
pages = {33107-33116},
year = {2020},
doi = {10.1073/pnas.2010827117},
URL = {https://www.pnas.org/doi/abs/10.1073/pnas.2010827117},
}

@Book{Milburn/Quantum_optics,
author = {Walls, D. F. and Milburn, G. J. },
title = { Quantum Optics},
edition = { Springer study ed. },
isbn = { 3540588310 },
publisher = { Springer-Verlag Berlin ; New York },
pages = { xii, 351 p. : },
year = { 1995 },
type = { Book },
language = { English },
subjects = { Quantum optics. },
life-dates = { 1995 -  },
catalogue-url = { https://nla.gov.au/nla.cat-vn674977 },
doi = {},
url = {}
}

@misc{Coecke/ZX/2023,
Author = {B. Coecke},
Title = {Basic ZX-calculus for students and professionals},
Year = {2023},
Eprint = {arXiv:2303.03163},
}

@misc{providers,
  author       = "Qiskit providers",
  title        = "https://qiskit.org/providers",
  howpublished = "",
  month        = "",
  year         = "2023",
}

@article{Cambell/dimmer/2013,
  title = {Dynamics of entanglement in a dissipative Bose-Hubbard dimer},
  author = {T. Pudlik and H. Hennig and D. Witthaut and D. K. Campbell},
  journal = {Phys. Rev. A},
  volume = {88},
  issue = {6},
  pages = {063606},
  numpages = {11},
  year = {2013},
  month = {Dec},
  publisher = {American Physical Society},
  doi = {10.1103/PhysRevA.88.063606},
  url = {https://link.aps.org/doi/10.1103/PhysRevA.88.063606}
}

@misc{Fujii/dissentangling/1999,
      title={A Universal Disentangling Formula for Coherent States of Perelomov's Type}, 
      author={K. Fujii and T. Suzuki},
      year={1999},
      eprint={hep-th/9907049},
      archivePrefix={arXiv},
      primaryClass={hep-th}
}

@article{Eckert/symmetrization/1997,
author = {A. Barenco and A. Berthiaume and D. Deutsch and A. Ekert and R. Jozsa and C. Macchiavello},
title = {Stabilization of Quantum Computations by Symmetrization},
journal = {SIAM Journal on Computing},
volume = {26},
number = {5},
pages = {1541-1557},
year = {1997},
doi = {10.1137/S0097539796302452},
URL = { https://doi.org/10.1137/S0097539796302452},
}

@Inbook{Moffat/Binary_Encoding/2008,
author="A. Moffat",
title="Compressing Integer Sequences and Sets",
bookTitle="Encyclopedia of Algorithms",
year="2008",
publisher="Springer US",
address="Boston, MA",
pages="178--183",
isbn="978-0-387-30162-4",
doi="10.1007/978-0-387-30162-4_84",
url="https://doi.org/10.1007/978-0-387-30162-4_84"
}

@article{Bouchard/HOM/2021,
doi = {10.1088/1361-6633/abcd7a},
url = {https://dx.doi.org/10.1088/1361-6633/abcd7a},
year = {2020},
month = {dec},
publisher = {IOP Publishing},
volume = {84},
number = {1},
pages = {012402},
author = {F. Bouchard and A. Sit and Y. Zhang and R. Fickler and F. M. Miatto and Y. Yao and F. Sciarrino and E. Karimi},
title = {Two-photon interference: the Hong–Ou–Mandel effect},
journal = {Reports on Progress in Physics},
}

@article{Hong/original_HOM/1987,
  title = {Measurement of subpicosecond time intervals between two photons by interference},
  author = {C. K. Hong and Z. Y. Ou and L. Mandel},
  journal = {Phys. Rev. Lett.},
  volume = {59},
  issue = {18},
  pages = {2044--2046},
  numpages = {0},
  year = {1987},
  month = {Nov},
  publisher = {American Physical Society},
  doi = {10.1103/PhysRevLett.59.2044},
  url = {https://link.aps.org/doi/10.1103/PhysRevLett.59.2044}
}

@article{Jin/spectral_HOM/2015,
author = {R. B. Jin and T. Gerrits and M. Fujiwara and R. Wakabayashi and T. Yamashita and S. Miki and H. Terai and R. Shimizu and M. Takeoka and M. Sasaki},
journal = {Opt. Express},
number = {22},
pages = {28836--28848},
publisher = {Optica Publishing Group},
title = {Spectrally resolved Hong-Ou-Mandel interference between independent photon sources},
volume = {23},
month = {Nov},
year = {2015},
url = {https://opg.optica.org/oe/abstract.cfm?URI=oe-23-22-28836},
doi = {10.1364/OE.23.028836},
}

@article{Omar/Entanglement_experiments22,
	Author = {Corona-Aquino, S. and Calder{\'o}n-Losada, O. and Li-G{\'o}mez, M. Y. and Cruz-Ramirez, H. and {\'A}lvarez-Venicio, V. and Carre{\'o}n-Castro, M. P. and de J. Le{\'o}n-Montiel, R. and U'Ren, A. B.},
	Booktitle = {The Journal of Physical Chemistry A},
	Da = {2022/04/14},
	Date = {2022/04/14},
	Doi = {10.1021/acs.jpca.2c00720},
	Isbn = {1089-5639},
	Journal = {The Journal of Physical Chemistry A},
	Journal1 = {J. Phys. Chem. A},
	M3 = {doi: 10.1021/acs.jpca.2c00720},
	Month = {04},
	Number = {14},
	Pages = {2185--2195},
	Publisher = {American Chemical Society},
	Title = {Experimental Study of the Validity of Entangled Two-Photon Absorption Measurements in Organic Compounds},
	Ty = {JOUR},
	Url = {https://doi.org/10.1021/acs.jpca.2c00720},
	Volume = {126},
	Year = {2022},
	Year1 = {2022}}

@article{Zelinger/BS_network/1994,
  title = {Experimental realization of any discrete unitary operator},
  author = {M. Reck and A. Zeilinger and H. J. Bernstein and P. Bertani},
  journal = {Phys. Rev. Lett.},
  volume = {73},
  issue = {1},
  pages = {58--61},
  numpages = {0},
  year = {1994},
  month = {Jul},
  publisher = {American Physical Society},
  doi = {10.1103/PhysRevLett.73.58},
  url = {https://link.aps.org/doi/10.1103/PhysRevLett.73.58}
}

@article{Bezerra/HOM_superselection_rules/2023,
doi = {10.1088/1367-2630/acfa1e},
url = {https://doi.org/10.1088/1367-2630/acfa1e},
year = {2023},
month = {sep},
publisher = {IOP Publishing},
volume = {25},
number = {9},
pages = {093047},
author = {Bezerra, M E O and Shchesnovich, V S},
title = {Families of bosonic suppression laws beyond the permutation symmetry principle},
journal = {New Journal of Physics}
}

@inproceedings{Aaronson/permanents/2011,
author = {S. Aaronson and A. Arkhipov},
title = {The Computational Complexity of Linear Optics},
year = {2011},
isbn = {9781450306911},
publisher = {Association for Computing Machinery},
address = {New York, NY, USA},
url = {https://doi.org/10.1145/1993636.1993682},
doi = {10.1145/1993636.1993682},
booktitle = {Proceedings of the Forty-Third Annual ACM Symposium on Theory of Computing},
pages = {333–342},
numpages = {10},
location = {San Jose, California, USA},
series = {STOC '11}
}

@misc{Scheel/permanents/2004,
      title={Permanents in linear optical networks}, 
      author={S. Scheel},
      year={2004},
      eprint={quant-ph/0406127},
      archivePrefix={arXiv},
      primaryClass={quant-ph}
}

@article{Mandel/PDC_experiments/1995,
author = {L. Mandel},
title = {Two-Photon Downconversion Experiments},
journal = {Ann. N. Y. Acad. Sci.},
volume = {755},
number = {1},
pages = {1-12},
doi = {https://doi.org/10.1111/j.1749-6632.1995.tb38952.x},
year = {1995}
}

@Article{Shaked/two_mode_quadrature/2018,
author={Y. Shaked
and Y. Michael
and R. Z. Vered
and L. Bello
and M. Rosenbluh
and A. Peer},
title={Lifting the bandwidth limit of optical homodyne measurement with broadband parametric amplification},
journal={Nature Communications},
year={2018},
month={Feb},
day={09},
volume={9},
number={1},
pages={609},
issn={2041-1723},
doi={10.1038/s41467-018-03083-5},
url={https://doi.org/10.1038/s41467-018-03083-5}
}

@misc{Biamonte/tensor_networks/2019,
Author = {J. Biamonte},
Title = {Lectures on Quantum Tensor Networks},
Year = {2019},
Eprint = {arXiv:1912.10049}
}
